\documentclass[11pt]{article}
\usepackage{amsfonts,amssymb,amsmath}

\def \RTITLE {{\small Harmonic Bilocal Fields}}
\newcommand{\thlabel}[1]{%
  \refstepcounter{Theorem}{\bf \arabic{section}.\arabic{Theorem}.\hspace{-1pt}}\label{#1}}
\newcommand{\dflabel}[1]{%
  \refstepcounter{Definition}{\bf \arabic{section}.\arabic{Definition}.\hspace{-1pt}}\label{#1}}

\newcounter{Theorem}

\newcounter{Definition}

\def \x {{\mathrm{x}}}
\def \y {{\mathrm{y}}}
\def \e {{\mathrm{e}}}
\def \u {{\mathrm{u}}}

\def \z {{\mathrm{z}}}
\def \w {{\mathrm{w}}}

\def \trn {^{\text{\rm tr}}}
\def \AA{{\mathbb A}}
\def \R {{\mathbb R}}      
\def \C {{\mathbb C}}      
\def \Z {{\mathbb Z}}      
\def \N {{\mathbb N}}      
\def \Sr {{\mathbb S}}     
\def \di {\partial}        
\def \spr {\cdot}          
\def \lvac {\langle 0 |}   
\def \rvac {|0\rangle}     
\def \FT {\mathcal{F}}     
\def \VS {\mathcal{V}}     
\def \AFE {\widehat{\mathcal{A}}}     
\def \AF {\mathcal{A}}     
\def \QQ {\mathcal{B}}
\def \VOL {\mathfrak{V}}   

\def \PROD {*}

\DeclareMathAlphabet{\mathrsfs}{U}{rsfs}{m}{n} 

\renewcommand{\geq}{\geqslant}
\renewcommand{\leq}{\leqslant}

\def \La {
\left\langle \!\!{\,}^{\mathop{}\limits_{}}_{\mathop{}\limits^{}}\right.}
\def \Ra {
\left. \!\!{\,}^{\mathop{}\limits_{}}_{\mathop{}\limits^{}}\right\rangle}
\def \Vl {
\left. \!\!{\,}^{\mathop{}\limits_{}}_{\mathop{}\limits^{}} \right|}

\newcommand{\beq}{\begin{equation}}
\newcommand{\eeq}{\end{equation}}
\newcommand{\beqa}{\begin{eqnarray}}
\newcommand{\eeqa}{\end{eqnarray}}
\newcommand{\nn}{\nonumber \\}

\def \setcntrs {\setcounter{equation}{0}\setcounter{Theorem}{0}\setcounter{Definition}{0}}

\newcounter{tmpc}
\newlength{\tmplenght}
\newlength{\tmplenghta}
\newlength{\tmplenghtb}
\newlength{\tmplenghtc}

\newenvironment{LIST}[1]{%
\setlength{\tmplenghta}{#1}
\setlength{\tmplenghtb}{#1}
\setlength{\tmplenghtc}{#1}
\advance\tmplenghtb-5pt
\advance\tmplenghtc 42pt
\setcounter{tmpc}{0}
\begin{list}{{\rm (\alph{tmpc})}}{\usecounter{tmpc}
\setlength{\leftmargin}{\tmplenghta}
\setlength{\rightmargin}{0cm}
\setlength{\itemsep}{1pt}
\setlength{\topsep}{3pt}
\setlength{\labelsep}{5pt}
\setlength{\labelwidth}{\tmplenghtb}
\setlength{\listparindent}{\tmplenghta}}
}{\end{list}}

\title{Harmonic bilocal fields generated by globally conformal
invariant scalar fields}

\author{Nikolay M.\ Nikolov$^{1,3}$, \\[2mm]
Karl-Henning Rehren$^{2}$, \\[2mm]
Ivan Todorov$^{1,3}$}

\begin{document}

\maketitle

\begin{center}
\scriptsize
\parbox{300pt}{
\begin{LIST}{21pt}
\item[$^{1}$]
Institute for Nuclear Research and Nuclear Energy, \\
Tsarigradsko Chaussee 72, BG-1784 Sofia, Bulgaria
\item[$^{2}$]
Institut f\"ur Theoretische Physik, Universit\"at G\"ottingen, \\
Friedrich-Hund-Platz 1, D-37077 G\"ottingen, Germany
\item[$^{3}$]
Abdus Salam International Centre for Theoretical Physics, \\
Strada Costiera 11, I--34014 Trieste, Italy
\end{LIST}
}
\end{center}

\begin{abstract}
\noindent
The twist two contribution in the operator product expansion of
$\phi_1 (\x_1)$ $\phi_2 (\x_2)$ for a pair of globally conformal invariant,
scalar fields of equal scaling dimension $d$ in four space--time dimensions
is a field $V_1 (\x_1,\x_2)$ which is harmonic in both variables.
It is demonstrated that the Huygens bilocality of $V_1$ can be equivalently
characterized by a ``single--pole property'' concerning the pole structure
of the (rational) correlation functions involving the product
$\phi_1(\x_1)$ $\phi_2 (\x_2)$.
This property is established for the dimension $d=2$ of $\phi_1$, $\phi_2$.
As an application we prove that any system of GCI
scalar fields of conformal dimension $2$ (in four space--time dimensions) can
be presented as a (possibly infinite) superposition of products of free
massless fields.
\end{abstract}




\normalsize

\section{Introduction}\label{Intr}
\setcntrs

Global Conformal Invariance (GCI) of Minkowski space Wightman fields
yields rationality of correlation functions \cite{NT01}.
This result opens the way for a nonperturbative construction and
analysis of GCI models for higher dimensional Quantum Field Theory
(QFT), by exploring further implications of the Wightman axioms. 

By choosing the axiomatic approach, we avoid any bias about
the possible origin of the model, because we aim at a broadest
possible perspective. On the other hand, the assumption of 
GCI limits the analysis to a class of theories that can be
parameterized by its (generating) field content and finitely many
coefficients for each correlation function (see Sect.~\ref{SE2}). 
As anomalous dimensions under the assumption of GCI are forced to
be integral, there is no perturbative approach {\em within} this
setting, but it is conceivable that a theory with a continuous
coupling parameter may exhibit GCI at discrete values (that appear as
renormalization group fixed points). An example of this type is
provided by the Thirring model: it is locally conformal invariant for
any value of the coupling constant $g$ and becomes GCI for positive
integer $g^2$ \cite{BMT88}.  

Previous axiomatic treatments of conformal QFT were focussed on
the representation theory and harmonic analysis of the conformal
group \cite{DMPPT,M77} as tools for the Operator Product Expansion
(OPE). The general {\em projective} realization of conformal symmetry
in QFT was already emphasized in \cite{SS74,SSV75} and found to
constitute a (partial) organization of the OPE. GCI is complementary in 
that it assumes true representations (trivial covering projection). 
A necessary condition for this highly symmetric situation is the 
presence of infinitely many conserved tensor currents (as we shall see
in Sect.~\ref{Sse3.3n}).  

The first cases studied under the assumption of GCI were theories
generated by a scalar field $\phi (\x)$ of (low) integral dimension
$d>1$. (The case $d=1$ corresponds to a free massless field with
a vanishing truncated $4$-point function $w_4\trn$.)
The cases $2\leqslant d \leqslant 4$,
which give rise to non-zero $w_4\trn$ were
considered in \cite{NST02,NST03,NRT05}.\footnote{%
The last two references are chiefly concerned with the case $d=4$
(in $D=4$ space-time dimensions) which appears to be of particular
interest as corresponding to a (gauge invariant) Lagrangian density.
The intermediate case $d=3$ is briefly surveyed in \cite{T06}.}

The main purpose in these papers was to study the constraints for
the $4$-point correlation (= Wightman) functions coming from the Wightman
(= Hilbert space) positivity.
This was achieved by using the conformal partial wave expansion.
An important technical tool in this expansion is the splitting of
the OPE into different twist contributions (see (\ref{e3.1})).
Each partial wave gives a nonrational contribution to the complete rational
$4$-point function. It is therefore remarkable that the sum of the leading,
twist two, conformal partial waves
(corresponding to the contributions of all conserved symmetric traceless
tensors in the OPE of basic fields) can be
proven in certain cases to be a rational function.
This means that the twist two part in the OPE of two fields $\phi$
is convergent in such cases to a bilocal field, $V_1(\x_1,\x_2)$,
which is our first main result in the present paper.
Throughout, ``bilocal'' means Huygens (= space--like {\em and} time--like)
locality with respect to both arguments. 
Proving bilocality exploits the bounds on the poles due to Wightman
positivity, and the conservation laws for twist two tensors which
imply that the bilocal fields are harmonic in both arguments. 

Trivial examples of harmonic bilocal fields are given by bilinear free
field constructions of the form $:\!\varphi(\x_1)\varphi(\x_2)\!:$,
$:\!\bar\psi(\x_1)\gamma_\mu(\x_1-\x_2)^\mu\psi(\x_2)\!:$, or 
$(\x_1-\x_2)^\mu(\x_1-\x_2)^\nu:\!F_{\mu\sigma}(\x_1)F^\sigma_\nu(\x_2)\!:$.
A major purpose of
this paper is to explore whether harmonic twist two fields can exist
which are not of this form, and whether they can be bilocal.
Moreover, we show that the presence of a bilocal field $V_1$ completely
determines the structure
of the theory in the case of a scaling dimension $d=2$.
The first step towards
the classification of $d=2$ GCI fields was made in \cite{NST02}
where the case of a unique scalar field was considered.
Here we extend our study to the most general case of a theory
generated by an arbitrary (countable) set of $d=2$ scalar fields. Our
second main result states that such fields are always combinations of
Wick products of free fields (and generalized free fields).  

\medskip

The paper is organized as follows.

Section~\ref{SE2} contains a review of relevant results concerning
the theory of GCI scalar fields.

In Sect.~\ref{Se4} we study conditions for the existence of the
harmonic bilocal field $V_1 (\x_1,\x_2)$.
We prove that Huygens bilocality of $V_1(\x_1,\x_2)$ is equivalent to
the \textit{single pole property} (SPP), Definition~\ref{D3.1}, which
is a condition on the pole structure of the leading singularities of
the truncated correlation functions of $\phi_1 (\x_1) \phi_2 (\x_2)$
whose twist expansion starts with $V_1(\x_1,\x_2)$. This nontrivial
condition qualifies a premature announcement in \cite{BNRT07} that
Huygens bilocality is automatic. 

Indeed, the SPP is trivially satisfied for all correlations of
free field constructions of harmonic fields with other (products of)
free fields, due to the bilinear structure of $V_1$. Thus any
violation of the SPP is a clear signal for a nontrivial field content of
the model. Moreover, the SPP will be proven from general principles
for an arbitrary system of $d=2$ scalar fields (the case studied
in \cite{BNRT07}). Yet, although the pole structure of $U(\x_1,\x_2)$ turns
out to be highly constrained in general by the conservation laws of twist
two tensor currents, the SPP does not follow for fields of higher dimensions,
as illustrated by a counter-example of a $6$-point function of $d=4$
scalar fields involving double poles (Sect.~\ref{Sse3.4}).

The existence of $V_1(\x_1,\x_2)$ as a Huygens bilocal field in a
theory of dimension $d=2$ fields allows to determine the truncated
correlation functions up to a single parameter in each of them. This
is exploited in Sect.~\ref{Se5}, where an associative algebra
structure of the OPE of $d=2$ scalar fields and harmonic bilocal
fields is revealed. The free-field representation of these fields is
inferred by solving an associated moment problem.

\section{Properties of GCI scalar fields}\label{SE2}
\setcntrs

\subsection{Structure of correlation functions and pole bounds}
\label{SE2-1}

We assume throughout the validity of the Wightman axioms for a QFT on
the $D=4$ flat Minkowski space--time $M$
(except for asymptotic completeness) -- see \cite{SW}.
Our results can be, in fact, generalized in a straightforward way
to any even space--time dimension $D$.
The condition of GCI \textit{in the Minkowski space}
is an additional symmetry condition on the correlation functions of
the theory \cite{NT01}. In the case of a scalar field $\phi (\x)$,
it asserts that the correlation functions of $\phi (\x)$ are
invariant under the substitution
\beq\label{e1.1}
\phi (\x) \, \mapsto \, \det\Bigl(\frac{\di g}{\di \x}\Bigr)^{\frac{d}{4}} \,
\phi \bigl(g(\x)\bigr)\;,
\eeq
where $\x \mapsto g(\x)$ is any conformal transformation of the
Min\-kow\-ski space, $\frac{\di g}{\di \x}$ is its Jacobi matrix and
$d > 0$ is the \textit{scaling dimension} of $\phi$.
An important point is that the invariance of Wightman functions
$\La 0 \Vl \phi (\x_1)$ $\cdots$ $\phi (\x_n) \Vl 0 \Ra$
under the transformation (\ref{e1.1}) should be
valid for all $\x_k \in M$ in the domain of definition of $g$
(in the sense of distributions).
It follows that $d$ must be an integer in order to ensure
the singlevaluedness of the prefactor
in (\ref{e1.1}).
Thus, GCI implies that only integral anomalous dimensions can occur.

\medskip

The most important consequen\-ces of GCI in the case of scalar fields
$\phi_k (\x)$ of dimensions $d_k$ are summarized as follows.

\medskip

(a)
\emph{Huygens Locality} (\cite[Theorem 4.1]{NT01}).
Fields commute for non light--like separations.
This has an algebraic version:
\beq\label{e2.1}
\bigl[(\x_1 - \x_2)^2\bigr]^{N}
\bigl[ \phi_1 (\x_1), \phi_2 (\x_2) \bigr] \, = \, 0
\eeq
for a sufficiently large integer $N$.

\medskip

(b)
\emph{Rationality of Correlation Functions} (cf.\ \cite[Theorem 3.1]{NT01}).
The general form of Wightman functions is:
\beq\label{e2.2}
\La 0 \Vl \phi_1 (\x_1) \cdots \phi_n (\x_n) \Vl 0 \Ra
\, = \, \mathop{\sum}\limits_{\{\mu_{jk}\}} \, C_{\{\mu_{jk}\}}
\, \mathop{\prod}\limits_{j < k} \,
(\rho_{jk})^{\mu_{jk}}\, ,
\eeq
where here and in what follows we set
\beq\label{rho}
\rho_{jk} \, := \, (\x_{jk} -i \, 0 \, \e_0)^2 = \, (\x_{jk})^2 +i \,
0 \, \x_{jk}^0
\, , \quad
\x_{jk} \, := \, \x_j - \x_k
\, ;
\eeq
the sum in Eq.~(\ref{e2.2}) is over all configurations of integral powers
$\{\mu_{jk}=\mu_{kj}\}$
subject to the following conditions:
\beq\label{e2.3}
\sum_{\quad j\,(\neq k)
} \mu_{jk}  = - d_k
, \ \
\eeq
and pole bounds
\(\mu_{jk} \geqslant -
\left[\!\!\left[\frac{\textstyle d_j + d_k}{\textstyle 2}
+ \frac{\textstyle \delta_{d_jd_k}-1}{\textstyle 2}\right]\!\!\right]\).
Equation (\ref{e2.3}) follows from
the conformal invariance under (\ref{e1.1});
the pole bounds express the absence of
non-unitary representations in the OPE of two fields \cite[Lemma
4.3]{NT01}. Under these conditions
the sum in (\ref{e2.2}) is always finite
and there are a finite number of free parameters for every
$n$-point correlation function. We shall refer to the form
(\ref{e2.2}) as a {\bf Laurent polynomial} in the variables
$\rho_{jk}$.%
\footnote{Writing correlation functions in terms of the
  conformally invariant cross ratios is particularly useful to
  parameterize $4$-point functions. A basis of cross ratios for an
  $n$-point function is used in the proof of Lemma \ref{L3.6}. The general systematics of the
  pole structure, however, is more transparent in terms of the present
  variables.}   
\medskip

(c)
The \textit{truncated} Wightman functions
$\La 0 \Vl \phi_1 (\x_1)$ $\cdots$ $\phi_n (\x_n) \Vl 0 \Ra\trn$
are of the same form like~(\ref{e2.2}) but with
pole degrees $\mu_{jk}\trn$ bounded by
\beq\label{e1.4n}
\mu_{jk}\trn > - \frac{d_j \!+ \! d_k}{2} \, 
\eeq
(cf. \cite[Corollary 4.4]{NT01}). 

The cluster condition, expressing the uniqueness of the vacuum,
requires that if a non-empty proper subset of points $\x_k$ among all
$\x_i$ ($i=1,\dots,n$) is shifted by 
$t\cdot{\mathrm a}$ $({\mathrm a}^2\neq 0)$, then the truncated 
function must vanish in the limit $t\to\infty$. For the two-point
clusters $\{\x_j,\x_k\}$, this condition is ensured by (\ref{e1.4n})
in combination with with (\ref{e2.3}). For higher clusters, it puts
further constraints on the admissible linear combinations of terms of
the form (\ref{e2.2}). Note however, that because of possible
cancellations the individual terms need not vanish in the cluster limit.  

The cluster condition will be used in establishing the single pole
property for $d=2$. 

\subsection{Twist expansion of the OPE and bi--harmonicity of twist
two contribution}\label{SE2-2}

The most powerful tool provided by GCI is the explicit construction of
the OPE of local fields in the general (axiomatic) framework.
\medskip

Let $\phi_1 (\x)$ and $\phi_2 (\x)$ be two GCI scalar fields of the
same scaling dimension $d$ and consider the operator distribution
\beq\label{e2.5}
U (\x_1,\x_2) \, = \, (\rho_{12})^{d-1} \,
\Bigl(\phi_1(\x_1)\,\phi_2(\x_2) -
\lvac\phi_1(\x_1)\,\phi_2(\x_2)\rvac\Bigr)\,.
\eeq
As a consequence of the pole bounds (\ref{e1.4n}), $U (\x_1,\x_2)$ is
\textit{smooth} in the difference $\x_{12}$.
This is to be understood in a weak sense for matrix elements of $U$
between bounded energy states. Obviously, $U (\x_1,\x_2)$ is a
\textbf{Huygens bilocal} field in the sense that
\beq\label{e2.6}
\hspace{0pt}
\bigl[(\x_1-\x)^2(\x_2-\x)^2\bigr]^N \bigl[U(\x_1,\x_2),\psi(\x)\bigr] = 0
\hspace{-3pt}
\eeq
for every field $\psi(\x)$ that is Huygens local with respect to $\phi_k(\x)$.
Then, one introduces the OPE of $\phi_1(\x_1)\,\phi_2(\x_2)$
by the Taylor expansion of $U$ in $\x_{12}$
\beq\label{e2.7}
U (\x_1,\x_2) = \mathop{\sum}_{n = 0}^{\infty} \
\mathop{\sum}_{\mu_1,\dots,\mu_n = 0}^{3} \
\x_{12}^{\mu_1} \cdots \x_{12}^{\mu_n} \, X_{\mu_1\dots\mu_n}^n(\x_2)\; ,
\eeq
where $X_{\mu_1\dots\mu_n}^n (\x_2)$ are Huygens local fields.
We can consider the series (\ref{e2.7}) as a formal power series,
or as a convergent series in terms of the analytically continued
correlation functions of $U (\x_1,\x_2)$. We will consider at this
point the series (\ref{e2.7}) just as a formal series. (See also
\cite{BN06} for the general case of constructing OPE via multilocal
fields in the context of vertex algebras in higher dimensions.)

\medskip

Since the prefactor in (\ref{e2.5}) transforms as a scalar density of
conformal weight $(1-d,1-d)$ then $U(\x_1,\x_2)$ transforms as a
conformal bilocal field of weight $(1,1)$. Hence,
the local fields $X_{\mu_1\dots\mu_n}^n$ in (\ref{e2.7})
have scaling dimensions $n+2$ but are not, in general,
quasiprimary.\footnote{%
Quasiprimary fields transform irreducibly under conformal transformations.}
One can pass to an expansion in quasiprimary fields by subtracting from
$X_{\mu_1\dots\mu_n}^n$ derivatives of lower dimensional fields
$X_{\mu_1\dots\mu_{n'}}^{n'}$.
The resulting quasiprimary fields $O_{\mu_1\dots\mu_{\ell}}^k$ are
traceless tensor fields of rank $\ell$ and dimension $k$.
The difference
\beq\label{e3.1}
k - \ell \quad
(\text{``dimension $-$ rank''})
\eeq
is called \textbf{twist} of the tensor field $O_{\mu_1\dots\mu_{\ell}}^k$.
Unitarity implies that the twist is non-negative \cite{M77}, and
by GCI, it should be an even integer.
In this way one can reorganize the OPE (\ref{e2.7}) as follows
\beq\label{e3.2}
U(\x_1,\x_2) = V_1(\x_1,\x_2) \, + \, \rho_{12} \,
V_2(\x_1,\x_2) + \, (\rho_{12})^2 \, V_3(\x_1,\x_2) +
\cdots \,,
\eeq
where $V_{\kappa}(\x_1,\x_2)$ is the part of the OPE (\ref{e2.7})
containing only twist $2\kappa$ contributions.
Note that Eq.~(\ref{e3.2}) contains also the information that
the twist $2\kappa$ contributions contain a factor
$(\rho_{12})^{\kappa-1}$ (i.e. $V_{\kappa}$ are ``regular''
at $\x_1=\x_2$), which is a nontrivial feature of this OPE (obtained
by considering $3$-point functions). Thus, the expansion in twists
can be viewed as a light-cone expansion of the OPE.

\medskip

Since the twist decomposition of the fields is conformally invariant
then each $V_{\kappa}$ will be behave, at least infinitesimally,
as a scalar $(\kappa,\kappa)$ density under conformal transformations.

\medskip

Every $V_{\kappa}$ is a complicated (formal) series in twist $2\kappa$
fields and their derivatives:
\beq\label{e3.3}
V_{\kappa} (\x_1,\x_2) \, = \,
\mathop{\sum}\limits_{\ell = 0}^{\infty}
\, K_\kappa^{\mu_1\dots\mu_{\ell}} (\x_{12},\di_{\x_2}) \,
O_{\mu_1\dots\mu_{\ell}}^{\ell + 2\kappa} (\x_2)\, ,
\eeq
where $K_\kappa^{\mu_1\dots\mu_{\ell}} (\x_{12},\di_{\x_2})$
are infinite formal power series in $\x_{12}$ with coefficients that are
differential operators in $\x_2$ acting on the quasiprimary fields $O$.
The important point here is that the series
$K_\kappa^{\mu_1\dots\mu_{\ell}} (\x_{12},$ $\di_{\x_2})$
can be fixed \textit{universally} for any (even generally) conformal QFT.
This is due to the universality of conformal $3$-point functions.
The explicit form of $K_\kappa^{\mu_1\dots\mu_{\ell}} (\x_{12},\di_{\x_2})$
can be found in \cite{DMPPT,DO01} (see also \cite{NST03}).

\medskip

Thus, we can at this point consider
$V_{\kappa} (\x_1,\x_2)$ only as generating series
for the twist $2\kappa$ contributions to the OPE
of $\phi (\x_1) \phi (\x_2)$ but we still do not know whe\-ther these
series would be convergent and even if they were, it would not be
evident whether they would give bilocal fields. In the next section we
will see that this is true for the leading, twist two part under
certain conditions, which are automatically fulfilled for $d=2$.
\medskip

The higher twist parts $V_{\kappa}$ ($\kappa > 1$) are certainly not
convergent to Huygens bilocal fields, since their $4$-point functions,
computed in \cite{NST03}, are not rational.

\medskip

The major difference between the twist two tensor fields and the higher
twist fields is that the former satisfy \textit{conservation laws}:
\beq\label{e4.1}
\di_{x_{\mu_1}} O_{\mu_1\dots\mu_{\ell}}^{\ell+2} (\x) \, = \, 0
\qquad (\ell \geqslant 1) \,.
\eeq
This is a well known consequence of the conformal invariance
of the $2$-point function and the Reeh--Schlieder theorem.
It includes, in particular, the conservation laws of the currents and
the stress--energy tensor.
It turns out that $V_1(\x_1,\x_2)$ encodes in a simple way this
infinite system of equations.

\medskip

\noindent
\textbf{Theorem}~\thlabel{T2.1} (\cite{NST03}) \
\textit{The system of differential equations (\ref{e4.1}) is equivalent to
the harmonicity of $V_1 (\x_1,\x_2)$ in both arguments
(\textbf{\emph{bi--harmon\-icity}}) as a formal series, i.e.,}
$$\Box_{\x_1} V_1 (\x_1,\x_2) = 0 = \Box_{\x_2} V_1 (\x_1,\x_2).$$
\medskip

The proof is based on the explicit knowledge of the $K$ series in
(\ref{e3.3}) and it is valid even if the
theory is invariant under infinitesimal conformal transformations only.

The separation of the twist two part in (\ref{e3.2}) amounts to a
splitting of $U$ of the form
\beq\label{e4.6uv}
U(\x_1,\x_2) \,=\, V_1(\x_1,\x_2) + \rho_{12}\,\widetilde{U}(\x_1,\x_2) \,.
\eeq
This splitting can be thought in terms of matrix elements of
$U(\x_1,\x_2)$ expanded as a formal power series according to
(\ref{e2.7}). It is unique by virtue of Theorem~\ref{T2.1}, due to the
following classical Lemma:
\medskip

\noindent
\textbf{Lemma}~\thlabel{L2.2} (\cite{BT77,BN06})
{\it Let $u (\x)$ be a formal power series in $\x \in \C^4$ (or, $\C^D$)
with coefficients in a vector space $V$. Then there exist unique formal
power series $v (\x)$ and $\widetilde{u} (\x)$ with coefficients in
$V$ such that
\beq\label{hd}
u (\x) = v (\x) + \x^2 \, \widetilde{u} (\x)
\eeq
and $v (\x)$ is harmonic in $\x$ (i.e., $\Box_{\x} \, v(\x) = 0$).
(\ref{hd}) is called the \textbf{\emph{harmonic decomposition}} of
$u(\x)$ (in the variable $\x$ around $\x=0$), and the formal power series
$v (\x)$ is said to be the \textbf{\emph{harmonic part}} of $u(\x)$.}

\section{Bilocality of twist two contribution to the OPE}\label{Se4}
\setcntrs

Let us sketch our strategy for studying bilocality of $V_1(\x_1,\x_2)$.

The existence of the field $V_1(\x_1,\x_2)$ can be
established by constructing its correlation functions.
On the other hand, every correlation function\footnote{This
  short-hand notation stands for $\La 0 \Vl \phi_3(\x_3) \cdots
  \phi_k(\x_k)$ $V_1(\x_1,\x_2)$ $\phi_{k+1}(\x_{k+1})$ $\cdots
  \phi_n(\x_n) \Vl 0 \Ra$, here and in the sequel.}
$\La \cdot V_1 (\x_1,\x_2) \cdot \Ra$ of $V_1$ is obtained (originally, as a
formal power series in $\x_{12}$) under the splitting (\ref{e4.6uv}).
It thus appears as a harmonic decomposition of the corresponding
correlation function $\La \cdot U (\x_1,\x_2) \cdot \Ra$ of $U$:
\beq\label{equ3.1x}
\La \cdot U (\x_1,\x_2) \cdot \Ra \, = \, \La \cdot V_1 (\x_1,\x_2) \cdot \Ra + \rho_{12} \,
\La \cdot \widetilde{U} (\x_1,\x_2) \cdot \Ra \, .
\eeq
Note that we should initially treat the left hand side of
(\ref{equ3.1x}) also as a formal power series in $\x_{12}$ in order to
make the equality meaningful.
It is important that this series is always convergent as a Taylor
expansion of a rational function in a certain domain around $\x_1=\x_2$
in $M_{\C}^{\times 2}$, for the complexified Minkowski space
$M_{\C} = M + i M$, according to the standard analytic properties of
Wightman functions. We shall show in Sect.~\ref{SSec3.1} that this
implies the separate convergence of both terms in the right hand side
of (\ref{equ3.1x}). Hence, the key tool in constructing $V_1$ are the
harmonic decompositions
\beq\label{equ3.2x}
F (\x_1,\x_2) \, = \, H (\x_1,\x_2) + \rho_{12} \, \widetilde{F} (\x_1,\x_2)
\eeq
of functions $F (\x_1,\x_2)$ that are analytic in certain
neighbourhoods of the diagonal $\{\x_1=\x_2\}$.

Recall that $H$ in (\ref{equ3.2x}) is uniquely fixed
as the harmonic part of $F$ in $\x_{1}$ around $\x_2$, due to
Lemma~\ref{L2.2}. This is equivalent to the harmonicity $\Box_{\x_1}
\, H (\x_1,\x_2)$ $=$ $0$.
On the other hand, according to Theorem~\ref{T2.1} we have to consider
also the second harmonicity condition on $H$, $\Box_{\x_2} \, H
(\x_1,\x_2)$ $=$ $0$, i.e., $H$ is the harmonic part in $\x_2$
around $\x_1$. This leads to some ``integrability'' conditions for the
initial function $F (\x_1,\x_2)$, which we study in Sect.~\ref{Sse3.2}.

Next, to characterize the Huygens bilocality of $V_1$, we should have
rationality of its correlation functions $\La \cdot V_1 (\x_1,\x_2)
\cdot \Ra$, which is due to a straightforward extension of the
arguments of \cite[Theorem 3.1]{NT01}. 
But we have started with the correlation functions of $U$, which are
certainly rational. Hence, we should study another condition on $U$,
namely that its correlation functions have a \textit{rational} harmonic
decomposition. We show in Sect.~\ref{Sse3.3n} that this is equivalent
to a simple condition on the correlation functions of $U$, which we
call ``Single Pole Property'' (SPP).

In this way we establish in Sect.~\ref{Sse3.3} that $V_1$ always exists as
a Huygens bilocal field in the case of scalar fields of dimension $d=2$.
However, for higher scaling dimensions one cannot anymore expect that $V_1$
is Huygens bilocal in general.
This is illustrated by a counter-example, involving the $6$-point function
of a system of $d=4$ fields, given at the end of Sect.~\ref{Sse3.4}.

\subsection{Convergence of harmonic decompositions}\label{SSec3.1}

To analyze the existence of the harmonic decomposition of a convergent
Taylor series we use the complex integration techniques introduced in
\cite{BN06}.

Let $M_{\C} = M + i M$ be the complexification of Minkowski space, which
in this subsection is assumed to be $D$--dimensional, and
$E$ $=$ $\bigl\{ \x $ $:$ $(i\,x^0,$ $x^1,$ $\dots,$ $x^{D-1})$
$\in$ $\R^D \bigr\}$ its Euclidean real submanifold, and
$\Sr^{D-1}\subset E$ the unit sphere in $E$. We denote by $\|$$\cdot$$\|$
the Hilbert norm related to the fixed coordinates in $M_{\C}$:
$\|\x\|^2$ $:=$ $|\x^0|^2$ $+$ $\cdots$ $+$ $|\x^{D-1}|^2$.

Let us also introduce for any $r > 0$ a real compact submanifold $M_r$
of $M_{\C}$:
\beq\label{eq4.7}
M_r \, = \,
\bigl\{
\zeta \in M_{\C} : \zeta = r \, e^{i \theta} \w,\
\vartheta \in [0,\pi],\,
\w \in \Sr^{D-1}
\bigr\} \,
\eeq
(note that $\vartheta \in [\pi,2\pi]$ gives another parameterization of $M_r$).
Then there is an integral representation for the harmonic part of a
convergent Taylor series.

\medskip

\noindent
\textbf{Lemma}~\thlabel{L3.1} {\rm (cf. \cite[Sect.~3.3 and Appendix A]{BN06})}
{\it
Let $u (\x)$ be a complex formal power series that is absolutely convergent
in the ball $\|\x\| < r$, for some $r> 0$, to an analytic function $U (\x)$.
Then the harmonic part $v(\x)$ of $u(\x)$ (around $\x=0$), which is provided by
Lemma~\ref{L2.2}, is absolutely convergent for
\beq\label{equ3.5x}
|\x^2| + 2 \, r \, \| \x \| < r^2 .
\eeq
The analytic function $V (\x)$ that is the sum of the formal power series
$v (\x)$ has the following integral representation:
\beq\label{eq4.5x1}
V(\x) \, = \, \mathop{\int}\limits_{\hspace{-5pt}M_{r'}}
\frac{d^D \z \bigl|_{M_{r'}}}{\VOL_1} \
\frac{1 - \frac{\x^2}{\z^2}}{\bigl[(\z-\x)^2\bigr]^{\frac{D}{2}}} \ U(\z)
\, , \qquad \VOL_1 \, = \,
\mathop{\int}\limits_{\hspace{-5pt}M_1} d^D \z \bigl|_{M_{1}}
= i\pi\vert \Sr^{D-1}\vert,
\eeq
where $r' < r$,
$|\x^2|$ $+$ $2 \, r' \, \| \x \|$ $<$ $r'^2$,
and the (complex) integration measure $d^D \z \bigl|_{M_{r'}}$
is obtained by the restriction of the complex volume form $d^{D}\z$
($= dz^0 \wedge$ $\cdots$ $\wedge dz^{D-1}$) on $M_{\C}$ ($\cong \C^D$)
to the real $D$--dimensional submanifold $M_{r'}$ (\ref{eq4.7}), $r'>0$.}

\medskip

\noindent
\textit{Proof.}
Consider the Taylor expansion in $\x$ of the function
$\bigl(1$ $-$ $\frac{\x^2}{\z^2}\bigr)\Big/
\bigl[(\z$ $-$ $\x)^2\bigr]^{\frac{D}{2}}$
and write it in the form (cf. \cite[Sect.~3.3]{BN06})
\beq\label{eq4.10}
\frac{1 - \frac{\x^2}{\z^2}}{\bigl[(\z - \x)^2\bigr]^{\frac{D}{2}}}
= \mathop{\sum}\limits_{\ell = \, 0}^{\infty} \, (\z^2)^{-\frac{D}{2}-\ell}
H_{\ell} (\z,\x) , \qquad
H_{\ell} (\z,\x) =
\sum_{\mu} h_{\ell\mu} (\z) \, h_{\ell\mu}(\x) ,
\eeq
where
$\{h_{\ell\mu} (\u)\}$
is an orthonormal basis of harmonic homogeneous polynomials of degree
$\ell$ on the sphere $\Sr^{D-1}$.
This expansion is convergent for
\beq\label{eq4.14new}
\bigl| \x^2 \bigr| + 2 \, \bigl| \z \spr \x \bigr| < \bigl| \z^2 \bigr|
\eeq
since its left--hand side is related to the generating function for
$H_{\ell}$:
\beq\label{eq4.15new}
\frac{1-\lambda^2\,\x^2\,\y^2}
{(1-2\,\lambda\,\x \spr \y +\lambda^2\,\x^2\,\y^2)^{\frac{D}{2}}} \, = \,
\mathop{\sum}\limits_{\ell \, = \, 0}^{\infty} \
\lambda^{\ell} \, H_{\ell} (\x,\y) \,,
\eeq
the expansion (\ref{eq4.15new}) being convergent for $\lambda \leqslant 1$ if
$|\x^2 \y^2| + 2 |\x \spr \y| < 1$.
Then if we fix $r' < r$ and $\z$ varies on $M_{r'}$, a sufficient
condition for (\ref{eq4.14new}) is
$|\x^2|$ $+$ $2 \, r' \, \| \x \|$ $<$ $r'^2$ (since
$\mathop{\sup}\limits_{\w \, \in \, \Sr^{D-1}} |\w \spr \x| = \| \x \|$).

On the other hand, writing $u (\z)$ $=$ $\sum_{k=0}^\infty$ $u_k(\z)$,
where $u_k$ are homogeneous polynomials of degree $k$, we get by the
absolute convergence of $u (\z)$ the relation
(valid for $|\x^2|$ $+$ $2 \, r' \, \| \x \|$ $<$ $r'^2$)
\beq\label{equ3.9x}
\mathop{\int}\limits_{\hspace{-5pt}M_{r'}}
\frac{d^D \z \bigl|_{M_{r'}}}{\VOL_1} \
\frac{1 - \frac{\x^2}{\z^2}}{\bigl[(\z-\x)^2\bigr]^{\frac{D}{2}}} \ U(\z)
= \mathop{\sum}\limits_{k,\ell \, = \, 0}^{\infty} \ \
\mathop{\int}\limits_{\hspace{-5pt}M_{r'}}
\frac{d^D \z \bigl|_{M_{r'}}}{\VOL_1} \ (\z^2)^{-\frac{D}{2}-\ell} \,
H_{\ell} (\x,\z) \, u_k (\z) \,.
\eeq
Noting next that in the parameterization (\ref{eq4.7}) of $M_{r'}$ we have
$d^D \z \bigl|_{M_{r'}}$ $=$
$i$ $r'{}^D$ $e^{i \, D \, \vartheta}$ $d\vartheta$ $\wedge$ $d\sigma (\w)$,
where $d\sigma (\w)$ is the volume form on the unit sphere, we obtain for
the right hand side of (\ref{equ3.9x}):
\beqa
\mathop{\sum}\limits_{k,\ell \, = \, 0}^{\infty} \quad
\mathop{\int}\limits_{\hspace{-7pt}0}^{\hspace{7pt}\pi} \!
\frac{d\vartheta}{i\pi} \, e^{i \vartheta (k-\ell)} \,
\mathop{\int}\limits_{\hspace{-5pt}\Sr^{D-1}}
\frac{d\sigma (\w)}{|\Sr^{D-1}|} \ H_{\ell} (\x,\w) \, u_k (\w) \,.
\nonumber
\eeqa
Now if we write, according to Lemma~\ref{L2.2},
$u_k (\z) = \mathop{\sum}\limits_{2j \, \leqslant \, k}$
$\mathop{\sum}\limits_{\mu'}$
$c_{k,j,\mu'}$ $(\z^2)^{j}$ $h_{k-2j,\mu'} (\z)$
then we get by the orthonormality of $h_{\ell, \mu} (\w)$
\beqa
\mathop{\sum}\limits_{k,\ell \, = \, 0}^{\infty} \
\mathop{\sum}\limits_{2j \, \leqslant \, k} \
\mathop{\sum}\limits_{\mu} \
\delta_{\ell, k-2j} \
\mathop{\int}\limits_{\hspace{-7pt}0}^{\hspace{7pt}\pi} \!
\frac{d\vartheta}{i\pi} \ e^{i \vartheta (k-\ell)} \,
c_{k,j,\mu} \, h_{k-2j,\mu} (\x) \nn
\qquad
\, = \,
\mathop{\sum}\limits_{k \, = \, 0}^{\infty} \ \mathop{\sum}\limits_{\mu} \
c_{k,0,\mu} \, h_{k,\mu} (\x) \, = \, v (\x) \, .
\nonumber
\eeqa
The latter proves both: the convergence of $v (\x)$ in the domain
(\ref{equ3.5x}) (since $r' < r$ was arbitrary) and the integral
representation (\ref{eq4.5x1}).
\hfill $\quad\Box$

\medskip

As an application of this result we will prove now

\medskip

\noindent
\textbf{Proposition}~\thlabel{P3.2}
{\it
For all $n$ and $k$, and for all local fields $\phi_j$ ($j=3,\dots,n$)
the Taylor series
\beq\label{eq4.5}
\La 0 \Vl \phi_3 (\x_3) \cdots \phi_k (\x_k)\;
V_1 (\x_1,\x_2)\; \phi_{k+1} (\x_{k+1}) \cdots
\phi_n (\x_n)\Vl 0 \Ra
\eeq
in $\x_{12}$ converge absolutely in the domain
\beq\label{eq4.6a}
\Bigl(\|\x_{12}\|+\sqrt{\|\x_{12}\|^2+\bigl|\x_{12}^2\bigr|}\Bigr)
\Bigl(\|\x_{2j}\|+\sqrt{\|\x_{2j}\|^2+\bigl|\x_{2j}^2\bigr|}\Bigr)
< \bigl|\x_{2j}^2\bigr| \quad \forall\; j
\eeq
($j = 3,\dots,n$).
They all are real analytic and independent of $k$ for mutually
nonisotropic points.}

\medskip

\noindent
\textit{Proof.}
Let
\beqa\label{eq4.10new}
&&
F_k (\x_{12},\x_{23},\dots,\x_{2n})
\nn && \quad
=
\La 0 \Vl \phi_3 (\x_3) \cdots \phi_k (\x_k)\;
U (\x_1,\x_2)\; \phi_{k+1} (\x_{k+1}) \cdots
\phi_n (\x_n)\Vl 0 \Ra \qquad
\eeqa
be the correlation functions, analytically continued in $\x_{12}$.

As $F_k$, which is a rational function, depends on $\x$ $:=$ $\x_{12}$
via a sum of products of powers $\bigl[(\x-\x_{2j})^2\bigr]^{-\mu_j}$
it has a convergent expansion in $\x$ for
\beq\label{eq4.12new}
\bigl| \x^2 \bigr| + 2 \, \bigl| \x \spr \x_{2j} \bigr|
< \bigl| \x_{2j}^2 \bigr| \,.
\eeq
If we want $F_k$ to have a convergent Taylor expansion for $\|\x\| < r$
we get the following sufficient condition
\beq\label{eq4.13new}
r^2 < \bigl| \x_{2j}^2 \bigr| - 2 \, r \, \| \x_{2j} \|
.
\eeq
By Lemma~\ref{L3.1} we conclude that the series
(\ref{eq4.5}) is convergent for
\beq\label{eq4.16new}
|\x_{12}^2| + 2 \, r \, \| \x_{12} \| < r^2 .
\eeq
Combining both (sufficient) conditions (\ref{eq4.13new}) and
(\ref{eq4.16new}) for $r$ we find that they are compatible if
$\|\x_{12}\|+\sqrt{\|\x_{12}\|^2+\bigl|\x_{12}^2\bigr|}
< \sqrt{\|\x_{2j}\|^2+\bigl|\x_{2j}^2\bigr|} - \|\x_{2j}\|$,
which is equivalent to (\ref{eq4.6a}).
\hfill $\quad\Box$

\medskip

Note that one can also prove a similar convergence property
for the correlation functions of several $V_1$.

\medskip

\noindent
\textit{Remark 3.1.}
The domain of convergence of (\ref{eq4.5}) should be Lorentz invariant.
Hence, (\ref{eq4.5}) are convergent in the smallest Lorentz invariant set
containing the domain (\ref{eq4.6a}).
Such a set is determined by the values of the invariants
$\x_{12}^2$, $\x_{2j}^2$ and $\x_{12} \spr \x_{2j}$
and it turns out to be the set
\beqa\label{eq4.6}
\bigl|\x_{12}^2\bigr|^{\frac{1}{2}}
\bigl|\x_{2j}^2\bigr|^{\frac{1}{2}} \leqslant \bigl|\x_{12} \spr \x_{2j}\bigr|
< \frac{\bigl(\bigl|\x_{2j}^2\bigr|^{\frac{1}{2}}
-\bigl|\x_{12}^2\bigr|^{\frac{1}{2}}\bigr)^2}{4} \quad\quad\nn
\text{or equivalently} \quad
\sqrt{\bigl|\x_{12}^2\bigr| \bigl|\x_{2j}^2\bigr| +
\bigl|\x_{12} \spr \x_{2j}\bigr|^2}
< \frac{\bigl(\bigl|\x_{2j}^2\bigr|^{\frac{1}{2}}-
\bigl|\x_{12}^2\bigr|^{\frac{1}{2}}\bigr)^2}{4} .
\eeqa
\smallskip

Outside the domain of convergence (\ref{eq4.6}), the correlations of
$V_1(\x_1,\x_2)$ have to be defined by analytic continuation. When the
correlations are rational, $V_1$ is Huygens bilocal, but the
counter-example presented in Sect.~\ref{Sse3.4} shows that rationality
is not automatic. Then, it is not even obvious that the continuations
are single--valued within the tube of analyticity required 
by the spectrum condition, i.e., that $V_1$ exists as a distribution
in all of $M\times M$. Nontrivial case studies, however, show that at
least for $\x_k$ space--like to both $\x_1$ and $\x_2$, the
continuation is single--valued and preserves the independence on the
position $k$ in (\ref{eq4.5}) where $V_1(\x_1,\x_2)$ is inserted. This
leads us to conjecture   
\medskip 

\noindent
\textbf{Conjecture}~\thlabel{Conj}
{\it The twist two field $V_1(\x_1,\x_2)$, whose correlations are
  defined as the analytic continuations of the harmonic parts of those
  of $U(\x_1,\x_2)$, exists and is bilocal in the ordinary sense, i.e., it
  commutes with $\phi(\x)$ and $V_1(\x,\x')$ if $\x$ and $\x'$ are
  space--like to $\x_1$ and $\x_2$.}    
\medskip

We hope to return to this conjecture elsewhere (see also the Note
added in proof). Note that the
argument that locality implies Huygens locality \cite{NT01} does
not pass to bilocal fields. 

\subsection{Consequences of bi--harmonicity}
\label{Sse3.2}

Now our objective is to find the harmonic decomposition of the rational
functions $F(\x_1,\x_2)$ that depend on $\x_1$ and $\x_2$ through the
intervals $\rho_{ik}$ $=$ $(\x_i-\x_k)^2$, $i=1,2$, $k=3,\dots,n$,
for some additional points $\x_3,\dots,\x_n$.
The $F$'s, as correlation functions of $U (\x_1,\x_2)$,
have the form
\beqa\label{equ3.17xx}
F (\x_1,\x_2) &=&
\sum_{q \, = \, 0}^M \ (\rho_{12})^{q} \
F_{q}(\x_1,\x_2) \equiv
\sum_{q \, = \, 0}^M \ (\rho_{12})^{q} \
F_{q}\Bigl( \{\rho_{ik}\}_{\{i,k\}\neq\{1,2\}} \Bigr), \qquad
\\ \label{equ3.17x}
F_{q}(\x_1,\x_2)
&=& \!\! \mathop{\sum}\limits_{\{\mu_{1i}\},\{\mu_{2i}\}} \!
C_{q,\{\mu_{1j}\},\{\mu_{2j}\}} \,
\mathop{\prod}\limits_{j=3}^n (\rho_{1j})^{\mu_{1j}}
\mathop{\prod}\limits_{j=3}^n (\rho_{2i})^{\mu_{2j}}\, ,
\hspace{20pt}
\eeqa
where $M \in \N$ and $\mu_{1j}$, $\mu_{2j}$ ($j=3,\dots,n$) are
integers $> -d$
such that
$\sum_{j \geqslant 3} \mu_{1j}$ $=$ $\sum_{j \geqslant 3}$ $\mu_{2j}=-1-q$,
and the coefficients $C_{q,\{\mu_{1j}\},\{\mu_{2j}\}} \,$ may depend on
$\rho_{jk}$ ($j,k\geq 3$).

If $H$ is the harmonic part of $F$ in $\x_{12}$, then the leading part
$F_0$ (of order $(\rho_{12})^0$) is also the leading part of $H$.
We shall now proceed to show that bi--harmonicity of $H$
(Theorem~\ref{T2.1}), together with the first principles of QFT
including GCI, implies strong constraints on $F_0$.

\medskip

\noindent
\textbf{Proposition}~\thlabel{P3.4}
\textit{Let $F_0(\x_1,\x_2)$ be as in (\ref{equ3.17x}), and let
  $H(\x_1,\x_2)$ be its harmonic part with respect to $\x_1$ around
  $\x_2$. Then $H$ is also harmonic with respect to $\x_2$, if and only if
  $F_0$ satisfies the differential equation
\beq\label{eq.ed} (E_1D_2 - E_2D_1)F_0=0,
\eeq
where
$E_1=\sum_{i=3}^n\rho_{2i}\partial_{1i}$ (with
$\partial_{jk}=\partial_{kj}=\frac{\partial}{\partial\rho_{jk}}$),
$D_1 = \sum_{3\leq j<k\leq n}\rho_{jk}\partial_{1j}\partial_{1k}$,
and similarly for $E_2$ and $D_2$, exchanging $1\leftrightarrow 2$.}

\medskip

\noindent
\textit{Proof.}
By Proposition~\ref{P3.2} (see also Remark 3.1) we can consider $H$
as a {\em function} in the $2n-3$ variables $\rho_{1i}$, $\rho_{2i}$
($i\geq 3$) and $\rho_{12}$, analytic in some domain
that includes $\rho_{12} = 0$.

Expanding $H=\sum_q(\rho_{12})^q H_q/q!$, the
functions $H_q$ are
homogeneous of degree $-1-q$ in both sets of variables $\rho_{1i}$ and
$\rho_{2i}$, and $H_0=F_0$. To impose the harmonicity with respect to the
variable $\x_1$, we use the identity \cite[App.\ C]{NRT05}
\beq\label{laprho}
\Box_{\x_1} F = -4 \Bigl[\mathop{\sum}\limits_{ 2\leq i<j \leq n} \,
\rho_{ij} \,
\partial_{{1i}} \partial_{{1j}} \, F \Bigr]\Bigr|_{\rho_{ij} \, =
 \, (\x_i-\x_j)^2} ,
\eeq
valid for homogeneous functions of $\rho_{1i}$ of degree $-1$, to
express the wave operator $\Box_{\x_1}$ as a differential operator
with respect to the set of variables $\rho_{1i}$ ($i \geq 2$). This yields
the recursive system of differential equations
\beq\label{rec1}
E_1H_{q+1}=-D_1H_q.
\eeq
Performing the same steps with respect to the variable $\x_2$, one obtains
\beq\label{rec1-1}
E_2H_{q+1}=-D_2H_q.
\eeq
Eq.\ (\ref{eq.ed}) then arises as the integrability condition for the
pair of inhomogeneous differential equations for $H_1$ (putting
$q=0$), observing that
$E_2E_1-E_1E_2 = \sum\rho_{1i}\partial_{1i} - \sum\rho_{2i}\partial_{2i}$
vanishes on $H_1$ by homogeneity.

Conversely, if (\ref{eq.ed}) is fulfilled, then $H_1$ exists and satisfies
$(D_1E_2-D_2E_1)H_1=-(D_1D_2-D_2D_1)H_0=0$ because $D_1$ and $D_2$
commute. But this is equivalent to $(D_2E_1-D_1E_2)H_1=0$, which is in
turn the
integrability condition for the existence of $H_2$, and so on.
It follows that bi--harmonicity imposes no further conditions on
the leading function $H_0=F_0$.
\hfill $\quad\Box$

\medskip

The differential equation (\ref{eq.ed}) imposes the following
constraints on the leading part $F_0$ of the rational correlation
function $F$ (\ref{equ3.17xx}):
\medskip

\noindent
\textbf{Corollary}~\thlabel{C3.5}
\textit{Assume that the function $F_0$ as in
(\ref{equ3.17x}) satisfies the differential equation (\ref{eq.ed}). Then} \\
(i) \textit{If $F_0$ contains a ``double pole'' of the form
$(\rho_{1i})^{\mu_{1i}}(\rho_{1j})^{\mu_{1j}}$ with $i\neq j$ and
$\mu_{1i}$ and $\mu_{1j}$ both negative, then its coefficients must be
regular in $\rho_{2k}$ ($k\neq i,j$).} \\
(ii) \textit{$F_0$ cannot contain a ``triple pole'' of the form
$(\rho_{1i})^{\mu_{1i}}(\rho_{1j})^{\mu_{1j}}(\rho_{1k})^{\mu_{1k}}$ with
$i,j,k$ all different and $\mu_{1i}$, $\mu_{1j}$, $\mu_{1k}$ all negative.
\\
The same hold true, exchanging $1\leftrightarrow 2$.}

\medskip

\noindent
\textit{Proof.}
Pick any variable, say $\rho_{2k}$, and decompose
$F_0=\sum_{r\geq -p}(\rho_{2k})^rf_r$ as a Laurent polynomial in $\rho_{2k}$.
The differential equation (\ref{eq.ed}) turns into the recursive system
$$
\left(\rho_{1k}\sum_{i<j}\rho_{ij}\partial_{1i}\partial_{1j}
- \sum_{i,j\neq k}\rho_{2i}\rho_{kj}\partial_{1i}\partial_{2j}\right) r
\cdot f_r = X_r f_{r-1} + Y f_r
$$
of differential equations for the functions $f_r$ which are Laurent
polynomials in the remaining variables. The precise form of the
polynomial differential operators $X_r$ and $Y$ does not
matter. Assume the lowest power $-p$ of $\rho_{2k}$ to be
negative. For $r=-p$, the right-hand-side vanishes. Because the term
$\rho_{ij}\partial_{1i}\partial_{1j}$ on the left-hand-side would produce a
singularity that cannot be cancelled by any other term, $f_{-p}$
cannot have a ``double pole'' in any pair of variables
$\rho_{1i},\rho_{1j}$ with $i\neq j$ and $i,j\neq k$. This property
passes recursively to all $f_r$ with $r<0$, because also the
right-hand-side never can contain such a pole.
This implies that a double pole in a pair of variables
$\rho_{1i},\rho_{1j}$ with $i\neq j$ cannot multiply a term that is
singular in $\rho_{2k}$ unless $k=i$ or $k=j$, proving~(i).

If the coefficient of the double pole were singular in $\rho_{1k}$,
$k\neq i,j$, then the resulting double pole in the pair $\rho_{1i}$,
$\rho_{1k}$ resp.\ $\rho_{1j}$, $\rho_{1k}$ would imply regularity
also in $\rho_{2j}$ resp.\ $\rho_{2i}$. Hence the coefficient of a
triple pole must be regular in all $\rho_{2m}$, which contradicts
the total homogeneity $-1$ of $F_0$ in these variables. This proves the
statement (ii).
\hfill $\quad\Box$

\subsection{A necessary and sufficient condition for Huygens bilocality}
\label{Sse3.3n}

\noindent
\textbf{Definition}~\dflabel{D3.1} (``Single Pole Property'', SPP)
{\it Let $f(\x_1,\ldots,\x_n)$ be a Laurent polynomial in the
  variables $\rho_{ij}$, i.e., regarded as a function of $\x_1$ only,
  it is a finite linear combination of functions of the form
\beq\label{e5.8nn}
\mathop{\prod}\limits_{j\geq 2} \,(\rho_{1j})^{\mu_{1j}}\equiv
\mathop{\prod}\limits_{j\geq 2} \,
\bigl[(\x_1-\x_j)^2\bigr]^{\mu_{1j}} ,
\eeq
where $\mu_{1j}$ ($j\geq 2$) are integers and the coefficients
may depend on the parameters $\rho_{jk}$ ($j,k\geq 2$). Then $f$
is said to satisfy the single pole property with respect to $\x_1$ if
it contains no terms for which there are $j\neq k$ ($j,k\geq 2$) such
that both $\mu_{1,j}$ and $\mu_{1,k}$ are negative. }
\medskip

The significance of SPP stems from the fact that
the harmonic parts $H$ of $F_0$, i.e., the
correlation functions of $V_1$, are again Laurent polynomials if and
only if $F_0$ satisfies the SPP. Namely, if $H$ is a harmonic Laurent
polynomial, the same argument as in \cite[Lemma C.1]{NRT05} (using the
representation (\ref{laprho}) of the wave operator) shows that $H$
fulfils the SPP with respect to $\x_1$, and so does $F_0$, because it
is the leading part of order $(\rho_{12})^0$ of $H$.
The converse is an immediate consequence of Lemma~\ref{L3.6}
(allowing for a relabelling and multiple counting of the points
$\x_3,\dots,\x_n$, which are not required to be distinct).

\medskip

\noindent
\textbf{Lemma}~\thlabel{L3.6}
\textit{Let $n\geq 4$. Every finite linear combination
of monomials of the form
\beq\label{g}
g_n(\x_1) = \frac{\prod_{i=4}^{n}\rho_{1i}}{(\rho_{13})^{n-2}}
\equiv \frac{\prod_{i=4}^{n}(\x_1-\x_i)^2}{[(\x_1-\x_3)^2]^{n-2}}
\eeq
has a \emph{rational} harmonic decomposition in $\x_1$ around $\x_2$
\beq\label{hdec}
g_n(\x_1) = h_n(\x_1) + (\x_1-\x_2)^2\cdot \tilde g_n(\x_1)
\eeq
i.e., $h_n$ is harmonic with respect to $\x_1$ and $\tilde g_n$ is regular
at $\x_1=\x_2$, and both $h_n$ and $\tilde g_n$ are rational. More
precisely, $(\rho_{13})^{n-2}(\rho_{23})^{n-3}h_n$ is a homogeneous
polynomial of total degree $2(n-3)$ in the variables $\{\rho_{ij}:
1\leq i < j\}$, which is separately homogoneous of degree $n-3$ in the
variables $\{\rho_{1i}: i\geq 2\}$ and in the variables $\{\rho_{12},
\rho_{2i}: i\geq 3\}$.}

\medskip

\noindent
\textit{Proof.} It is convenient to introduce the variables
\beq\label{stu}
t_i = \frac{\rho_{1i}\rho_{23}}{\rho_{13}\rho_{2i}}, \qquad
s_i = \frac{\rho_{12}\rho_{3i}}{\rho_{13}\rho_{2i}}, \qquad
u_{ij} = \frac{\rho_{12}\rho_{23}\rho_{ij}}
{\rho_{13}\rho_{2i}\rho_{2j}} \quad (4\leq i<j\leq n).
\eeq
We claim that $h_n(\x_1)$ is of the form
\beq\label{harm}
h_n(\x_1) =
\left(\prod_{i=4}^{n}\frac{\rho_{2i}}{\rho_{23}}\right)\cdot\frac
{f_n(t_i,s_i,u_{ij})}{\rho_{13}} \, ,
\eeq
where $f_n$ are polynomials of degree $n-3$ such that
$f_n(t_i,s_i=0,u_{ij}=0) = \prod_{i=4}^{n} t_i$. Because all $s_i$ and
$u_{ij}$ contain a factor $\rho_{12}$, these properties ensure that
$\tilde g_n$ given by $(g_n-h_n)/\rho_{12}$ is regular in
$\rho_{12}$.

Using again the identity (\ref{laprho}) for the wave operator, and
transforming this into a differential operator with respect to the set
of variables (\ref{stu}), we find
\beq\label{waveop}
\Box_{\x_1}\;h_n(\x_1) = -4
\left(\prod_{i=4}^{n}\frac{\rho_{2i}}{\rho_{23}}\right)
\frac{\rho_{23}}{(\rho_{13})^2\rho_{12}}\cdot Df_n(t_i,s_i,u_{ij}),
\eeq
where $D$ is the differential operator
\beq\label{D:stu}
D = (1+t\partial_t + s\partial_s + u\partial_u)(s\partial_t +
s\partial_s + u\partial_u) - (s\partial_s+u\partial_u)\partial_t -
u\partial_t\partial_t
\eeq
with shorthand notations for degree-preserving operators
$$t\partial_t = \sum_{i=4}^{n} t_i\partial_{t_i},\quad s\partial_t =
\sum_{i=4}^{n} s_i\partial_{t_i},\quad s\partial_s = \sum_{i=4}^{n}
s_i\partial_{s_i},
\quad u\partial_u = \sum_{4\leq i<j\leq n}u_{ij}\partial_{u_{ij}}$$
and degree-lowering operators
$$
\partial_t = \sum_{i=4}^{n} \partial_{t_i},\qquad u\partial_t\partial_t =
\sum_{4\leq i<j\leq n}u_{ij}\partial_{t_i}\partial_{t_j}.
$$
To solve the condition $Df_n=0$ for harmonicity, we make an ansatz
$$f_n(t_i,s_i,u_{ij}) = \sum_{K\subset N}g_K^{(n)}(s_k,u_{kl})\cdot
\prod_{i\in N\setminus K}(t_i-s_i), $$
where $N\equiv\{4,\dots,n\}$, $g_K^{(n)}$ are polynomials in the
variables $s_k$, $u_{kl}$ ($k,l\in K$) only, and $g^{(n)}_\emptyset =
1$. Then the harmonicity condition $Df_n=0$ is equivalent to the
recursive system
$$
(n-2-\vert K\vert+\Delta)\Delta\; g^{(n)}_K =
\Delta \sum_{k\in K}g^{(n)}_{K\setminus\{k\}} +
\sum_{k,l\in K, k<l}(u_{kl}-s_k-s_l)\,g^{(n)}_{K\setminus\{k,l\}},
$$
where
$|K|$ is the number of elements of the set K and
the differential operator $\Delta = s\partial_s + u\partial_u$
measures the total polynomial degree $r$ in $s_k$ and $u_{kl}$. Since
one can divide by $(n-2-\vert K\vert+r)r$ if $r>0$, there is a unique
polynomial solution such that $g^{(n)}_K(s_k=0,u_{kl}=0)=0$
($K\neq\emptyset$), and $g^{(n)}_K$ is of order $\leq \vert
K\vert$. So $f_n$ is of order $n-3$. (Explicitly, the first three
functions are $f_3=1$, $f_4=t_4-s_4$ and
$f_5=(t_4-s_4)(t_5-s_5)+\frac12(u_{45}-s_4-s_5)$.)
An inspection of
the recursion also shows that all possible factors $\rho_{2i}$ in the
denominators of the arguments of $f_n$ cancel with the factors in the
prefactor in (\ref{harm}), thus $h_n$ can have poles only in
$\rho_{13}$ and $\rho_{23}$
of the specified maximal degree. This proves the Lemma.
\hfill $\quad\Box$
\medskip

The upshot of the previous discussion is a necessary and sufficient
condition for the Huygens bilocality of $V_1$ which directly refers to
the local correlation functions of the theory: 
\medskip

\noindent
\textbf{Theorem}~\thlabel{T3.7}
\textit{The field $V_1 (\x_1,$ $\x_2)$ weakly converges on bounded
  energy states to a Huygens bilocal field which is conformal of
  weight $(1,1)$, if and only if the leading parts $F_0$ of the
  Laurent polynomials $F$ (\ref{equ3.17xx}) satisfy
  the ``single pole property'' (Def.\ \ref{D3.1}) with respect to both
  $\x_1$ and $\x_2$. In this case, the formal series $H$ converge to
  Laurent polynomials in $(\x_i-\x_j)^2$ subject to the
  same pole bounds, specified in Theorem~\ref{T2.1}, as $F$.}

\medskip

\noindent
\textit{Proof.}
We know already that if $V_1$ is a Huygens bilocal field, then its correlation
functions $H$ are Laurent polynomials of the form (\ref{e2.2}), and
that this implies the SPP for $F_0$ with respect to $\x_1$ and
$\x_2$. Conversely, if the SPP holds for $F_0$ with respect to $\x_1$
and $\x_2$, then $H$ are Laurent polynomials by Lemma~\ref{L3.6}, and hence
$V_1$ is relatively Huygens bilocal with respect to the fields $\phi_i$.
Since the general argument \cite{B60} that relative locality implies local 
commutativity of a field with itself refers only to local fields,
we want to give an explicit argument for the case at hand.

All the previous remains true
when in (\ref{eq4.5}) or (\ref{equ3.17xx}) a product of fields
$\phi_k(\x_k)_{k+1}\phi(\x_{k+1})$ is replaced by $U(\x_k,\x_{k+1})$. By
assumption, and because $U$ is bilocal, the contributions of order
$(\rho_{k,k+1})^0$ to the correlation functions of $U(\x_k,\x_{k+1})$
fulfil the SPP with respect to $\x_k$ and $\x_{k+1}$. By Lemma~\ref{L3.6},
this property is preserved upon the passage to the harmonic parts with
respect to $\x_1$ and $\x_2$. One may therefore continue in the same
way with $\x_k$, $\x_{k+1}$, and eventually find that all mixed
correlation functions of $\phi$'s and $V_1$'s converge to rational
functions. By this convergence we conclude that all products of
$\phi$'s and $V_1$'s converge on the vacuum, and this then defines
$V_1$ as a Huygens bilocal field, since its matrix elements will
satisfy Huygens locality.

The conformal properties of $V_1$ follow from the preservation of the
homogeneity and the
pole degrees in the harmonic decomposition, as
guaranteed by Lemma~\ref{L3.6}.
\hfill $\quad\Box$
\medskip

For $n=4$ points, the SPP is trivially satisfied because of
homogeneity. Hence the $4$-point function $\La 0 \Vl V_1^*V_1 \Vl 0
\Ra$ is always rational. It follows that its expansion in
(transcendental) partial waves \cite{NRT05} cannot terminate. This
means that (unless $V_1=0$ in which case there is not even a
stress-energy tensor) {\em a GCI QFT necessarily contains infinitely
  many conserved tensor fields of arbitrarily high spin.}

\subsection{The case of dimension 2}\label{Sse3.3}

Let us consider now the case of scalar fields $\phi_k$ of dimension 2.
We claim that in this case, Corollary~\ref{C3.5} in combination with
the cluster condition is sufficient to establish the SPP,
Definition~\ref{D3.1}. Hence we conclude by Theorem~\ref{T3.7} that
the twist two harmonic fields $V_1(\x_1,\x_2)$ are indeed Huygens bilocal fields.

To prove our claim, we use that by (\ref{e1.4n}),
$\mu_{ij}\geq-1$, hence the SPP is equivalent to the statement that 
there can be no term contributing to $\La 0 \Vl \phi_1 (\x_1) \cdots
$ $\phi_n (\x_n) \Vl 0 \Ra$, for which there is $i$ with more 
than two $\mu_{ij}$ negative ($j\neq i$). Thus assume that  
there is a term with, say, $\mu_{12}=\mu_{13}=\mu_{14}=-1$. It
constitutes a double pole for each of the three harmonic fields
$V_1(\x_1,\x_j)$ ($j=2,3,4$). Then by homogeneity (\ref{e2.3}), there
must be more poles in $\x_j$ ($j=2,3,4$), but these cannot be of the
form $\rho_{jk}$ with $k>4$ by Corollary~\ref{C3.5}. Hence (up to
permutations of $2,3,4$) $\mu_{23}=\mu_{24}=-1$, $\mu_{34}=0$. 
Again by homogeneity (\ref{e2.3}), the dependence on $\x_1,\dots,\x_4$
must be given by a linear combination of terms 
\beq\label{contra}
\frac {\rho_{1k}\rho_{4\ell}}
{\rho_{12}\rho_{13}\rho_{14}\rho_{23}\rho_{24}}
\eeq
with $k,\ell>4$. Applying the cluster limit (Sect.~\ref{SE2-1}) to 
the points $\x_1,\x_2,\x_3,\x_4$ in (\ref{contra}), the limit diverges
$\sim t^4$. This behavior is tamed to $\sim t^2$ by
anti--symmetrization in $k,\ell$, but it cannot be cancelled by any 
other terms. Hence the assumption leads to a contradiction. 

This proves the SPP if the generating scalar fields have dimension
$d=2$.

\subsection{A $d=4$ $6$-point function violating the SPP}\label{Sse3.4}

We proceed with an example of $6$-point function violating the SPP
in the case of two $d=4$ GCI scalar fields $L_i(\x)$ such that the
bilocal field $U(\x_1,\x_2)$ obtained from $L_1(\x_1)L_2(\x_2)$ has a
non-zero skew--symmetric part. Let $L$ be any linear combination of
$L_1$ and $L_2$.

The following admissible contribution to the truncated part of the $6$-point
function $\lvac U (\x_1,\x_2)L(\x_3)L(\x_4)U (\x_5,\x_6)\rvac$
clearly violates the SPP:
\beq\label{equ3.30x}
F_0(\x_1,\x_2) =
  \AA_{12}\AA_{56}\left[\frac{\rho_{15}\rho_{26}\rho_{34} - 2
  \rho_{15}\rho_{23}\rho_{46} - 2
  \rho_{15}\rho_{24}\rho_{36}}{\rho_{13}\rho_{14}\rho_{23}\rho_{24}\cdot
  \rho_{34}\cdot \rho_{35}\rho_{45}\rho_{36}\rho_{46}}\right]\;,
\eeq
where $\AA_{ij}$ stands for the antisymmetrization in the arguments
$\x_i$, $\x_j$. It is admissible as a truncated $6$-point
structure because $(\rho_{12}\rho_{56})^{-3}F_0$ obeys all the pole
bounds of Sect.~\ref{SE2} for a correlation
$\lvac L_1(\x_1)L_2(\x_2)L(\x_3)L(\x_4)L_1(\x_5)$ $L_2(\x_6) \rvac\trn$
of six fields of dimension $d=4$.

On the other hand, $F_0$ satisfies the differential
equation
\beq \label{ed.ex}
(E_1 D_2-E_2 D_1)F_0(\x_1,\x_2)=0
\eeq
(and similar in the variables $\x_5$ and
$\x_6$), ensuring that $F_0$ is the leading part of a bi--harmonic
function, analytic in a neighborhood of $\x_1=\x_2$ and $\x_5=\x_6$,
representing a contribution to the twist two $6$-point function
$\lvac V_1(\x_1,\x_2)L(\x_3)L(\x_4)$ $V_1(\x_5,\x_6) \rvac$, of which
$F_0$ is the leading part. This function cannot be
a Laurent polynomial in the $\rho_{ij}$ by our general argument that
the leading part of a bi--harmonic Laurent polynomial cannot satisfy
the SPP. Hence the twist two field $V_1(\x_1,\x_2)$ cannot be
Huygens bilocal.

The resulting contribution to the conserved local current $4$-point
function $\lvac J_\mu(\x_1)L(\x_3)L(\x_4)J_\nu(\x_5)\rvac\trn$ is obtained
through
$J_\mu(\x) = i
(\partial_{\x^\mu}-\partial_{\y^{\mu}})$ $V_1(\x,\y)\vert_{\x=\y}$. It
also satisfies the pertinent pole bounds. This structure is rational
as it should, because only the leading part $F_0$ contributes.
In fact, while the $6$-point structure involving the harmonic field
cannot be reproduced by free fields due to its double pole,
the resulting $4$-point structure does arise as one of the three
independent connected structures contributing to $4$-point functions
involving two Dirac currents $:\!\bar\psi_a\gamma^\mu\psi_b\!:$ and two
Yukawa scalars $\varphi:\!\bar\psi_c\psi_d\!:$ (allowing for internal
flavours $a,b,\dots$).

\section{The theory of GCI scalar fields of scaling dimension $d=2$}
\label{Se5}
\setcntrs

The scaling dimension $d=2$ is the minimal dimension of a GCI scalar field
for which one could expect the existence of nonfree models.
It turns out however, that in this case the fields can be constructed
as composite fields of free, or generalized free, fields.
Namely, we will establish the following result.
\medskip

\noindent
\textbf{Theorem}~\thlabel{T4.1}
\textit{Let $\{\Phi_m (\x)\}_{m \, = \, 1}^{\infty}$ be
a system of real GCI scalar fields of scaling dimension $d = 2$.
Then it can be realized
by a system of generalized free fields $\{\psi_m (\x)\}$
and a system of independent real massless free fields $\{\varphi_m (\x)\}$,
acting on a possibly larger Hilbert space, as follows:
\beq\label{eqXX}
\Phi_m (\x) \, = \,
\mathop{\sum}\limits_{j \, = \, 1}^{\infty} \,
\alpha_{m,j} \, \psi_j(\x) \, + \,
\frac{1}{2} \,
\mathop{\sum}\limits_{j,k \, = \, 1}^{\infty} \, \beta_{m,j,k} :
 \! \varphi_j (\x) \varphi_k (\x) \! : \, , \quad
\eeq
\vskip-3mm\noindent
where $\alpha_{m,j}$ and $\beta_{m,j,k}=\beta_{m,k,j}$ are real constants
such that $\mathop{\sum}\limits_{j \,=\, 1}^{\infty} \alpha_{m,j}^2 < \infty$
and $\mathop{\sum}\limits_{j,k \,=\, 1}^{\infty} \beta_{m,j,k}^2 < \infty$.
Here, we assume the normalizations
$\lvac \, \varphi_j(\x_1)$ $\varphi_k(\x_2) \, \rvac
= \delta_{jk} \, (\rho_{12})^{-1}$,
$\lvac \, \psi_j (\x_1)$ $\psi_k (\x_2) \, \rvac$ $
= \delta_{jk} \, (\rho_{12})^{-2}$.}
\medskip

The proof of Theorem~\ref{T4.1} is given at the end of Sect.~\ref{Sse4.2nn}.
The main reason for this result is the fact that in the $d=2$ case the
harmonic bilocal fields exist and furthermore, they are Lie fields.
This was originally recognized in \cite{NST02}, \cite{BNRT07} under the
assumption that there is a unique field $\phi$ of dimension~$2$. We are
extending here the result to an arbitrary system of $d=2$ GCI scalar fields.

If we assume the existence of a stress-energy tensor as a Wightman
field\footnote{A stress-energy tensor always exists as a
quadratic form between states generated by the fields $\Phi_m$ from
the vacuum \cite{DR03}.}, the generalized free fields must be absent in
(\ref{eqXX}), and the number of free fields must be finite. In this
case, the iterated OPE generates in particular the bilocal field
$\frac 12 \sum_i{:}\varphi_i(\x)\varphi_i(\y){:}$. As this field has 
no other positive-energy representation than those occurring in the
Fock space \cite{BNRT07}, nontrivial possibilities for correlations
between non-free fields and the fields (\ref{eqXX})  
are strongly limited.

\subsection{Structure of the correlation functions}
\label{Sse4.1}

We consider a GCI QFT generated by a set of hermitean (real) scalar
fields.
We denote by $\FT$ the \textit{real} vector space of \textit{all}
GCI real scalar fields of scaling dimension $2$ in the theory.
(Note that the space $\FT$ may be larger than the linear span of
the original system of $d=2$ fields of Theorem~\ref{T4.1}.)
We shall find in this section the explicit form of
the correlation functions of the fields from $\FT$.
\medskip

\noindent
\textbf{Theorem}~\thlabel{T4.2}
{\it
Let $\phi_1(\x),$ $\dots,$ $\phi_n (\x)$ $\in$ $\FT$
then their \emph{truncated} $n$-point functions have the form
\beq\label{eq5.1b}
\lvac \phi_1(\x_1) \, \cdots \, \phi_n(\x_n) \rvac\trn
=
\frac{1}{2n} \,
\mathop{\sum}\limits_{\sigma \, \in \, \mathcal{S}_n}
c^{(n)} (\phi_{\sigma_1},\dots,\phi_{\sigma_n})
\,
\bigl(\rho_{\sigma_1\sigma_2} \cdots \rho_{\sigma_n\sigma_1}\bigr)^{-1} ,
\eeq
where $c^{(n)}$ are multilinear functionals $c^{(n)}:\FT^{\otimes n} \to \R$
with the inversion and cyclic symmetries $c^{(n)} (\phi_1,$ $\dots,$
$\phi_n)$ $=$ $c^{(n)} (\phi_n,\dots,\phi_1)$ $=$ 
$c^{(n)} (\phi_n,$ $\phi_1,$ $\dots,$ $\phi_{n-1})$.}

\medskip

Before we prove the theorem,
let us first illustrate it on the example of the free field
realization (\ref{eqXX}).
In this case one finds
\beqa\label{eq5.3nn}
c^{(2)} \bigl(\Phi_{m_1},\Phi_{m_2} \bigr) \, &=& 
\mathop{\sum}\limits_{j \, = \, 1}^{\infty} \alpha_{m_1,j} \alpha_{m_2,j} +
\mathop{\sum}\limits_{j,k \, = \, 1}^{\infty}
\beta_{m_1,j,k} \, \beta_{m_2,j,k}
\nn \, &\equiv&
\mathop{\sum}\limits_{j \, = \, 1}^{\infty} \alpha_{m_1,j} \alpha_{m_2,j} + \,
\text{Tr} \, \beta_{m_1} \beta_{m_2}
\,,\quad
\nn
c^{(n)} \bigl(\Phi_{m_1},\dots,\Phi_{m_n}\bigr) \, &=&
\text{Tr} \, \beta_{m_1} \cdots \beta_{m_n}
\quad \text{for} \quad n > 2
\eeqa
where $\beta_m = \bigl(\beta_{m,j,k}\bigr)_{j,k}$.


\medskip

\noindent
\textit{Proof of Theorem~\ref{T4.2}.} We first recall the general
form (\ref{e2.2}) of the truncated correlation function with pole 
bounds (\ref{e1.4n}) that read in this case: $\mu_{jk}\trn
\geqslant -1$. The argument in Sect.~\ref{Sse3.3} shows that the
nonzero contributing terms in Eq.~(\ref{e2.2}) have for every
$j=1,\dots,n$ exactly \textit{two} negative $\mu_{jk}\trn$ or
$\mu_{kj}\trn$ for some $k=k_1,k_2$ different from $j$. 

The nonzero terms are therefore products of ``disjoint cyclic 
products of propagators'' of the form 
$1/\rho_{k_1k_2}\rho_{k_2k_3}\cdots\rho_{k_{r-1}k_r}\rho_{k_rk_1}$. 
But cycles of length $r<n$ are in conflict with the cluster condition
(Sect.~\ref{SE2}). We conclude that $\lvac \phi_1(\x_1)$ $\cdots$
$\phi_n(\x_n) \rvac\trn$ is a linear combination of terms like those
in (\ref{eq5.1b}) with some coefficients $c_{\sigma} (\phi_1,$
$\dots,$ $\phi_n)$ depending on the permutations $\sigma \in
\mathcal{S}_n$ and on the fields $\phi_j$ (multilinearly).
Locality, i.e.\
$\lvac \phi_1(\x_1)$ $\cdots$ $\phi_n(\x_n) \rvac\trn$ $=$
$\lvac \phi_{\sigma_1'}(\x_{\sigma_1'})$ $\cdots$
$\phi_{\sigma_n'}(\x_{\sigma_n'}) \rvac\trn$,
then implies
$c_{\sigma'\sigma} (\phi_1,\dots,\phi_n)$ $=$
$c_{\sigma} (\phi_{\sigma'_1},\dots,\phi_{\sigma'_n})$
($\sigma,\sigma' \in \mathcal{S}_n$),
so that
$c_{\sigma} (\phi_1,$ $\dots,$ $\phi_n)$ $=$ $c^{(n)}
(\phi_{\sigma_1},\dots,\phi_{\sigma_n})$
for some $c^{(n)} : \FT^{\otimes n} \to \R$.
The equalities
$c^{(n)} (\phi_1,\dots,\phi_n)$ $=$
$c^{(n)} (\phi_n,\dots,\phi_1)$ $=$ $c^{(n)} (\phi_n,\phi_1,\dots,\phi_{n-1})$
are again due to locality.
\hfill $\quad\Box$

\medskip

As we already know by the general results of the previous section, 
the harmonic bilocal field exist in the case of fields of dimension $d=2$.
Moreover, the knowledge of the correlation functions of the $d=2$ fields
allows us to find the form of the correlation functions of the
resulting bilocal fields.
This yields an algebraic structure in the space of real
(local and bilocal) scalar fields, which we proceed to display.

\medskip

Let us introduce together with the space $\FT$ of $d=2$ fields also
the real vector space $\VS$ of all real harmonic bilocal fields.
We shall consider $\FT$ and $\VS$ as built starting from our original
system of $d=2$ fields $\{\Phi_m\}$ of Theorem~\ref{T4.1}, by the following
constructions.

\medskip

(a) If $\phi_1(\x), \phi_2 (\x) \in \FT$ then introducing the bilocal
$(1,1)$--field $U(\x_1,\x_2) = \x_{12}^2 \Bigl[\phi_1(\x_1) \phi_2(\x_2) -
\lvac \phi_1(\x_1)\phi_2(\x_2) \rvac \Bigr]$ in accord
with Eq.~(\ref{e2.5}), 
we consider its harmonic decomposition
$U (\x,\y)$ $=$ $V_1 (\x,\y)$ $+$ $(\x-\y)^2$ $\widetilde{U} (\x,\y)$.
We denote $V_1 (\x,\y)$ by $\phi_1 \PROD \phi_2$;
this defines a bilinear map
$\FT \otimes \FT \mathop{\to}\limits^{\PROD} \VS$.

\medskip

(b) If now $v (\x,\y) \in \VS$ then $v^t (\x,\y)$ $:=$ $v (\y,\x)$ also
belongs to $\VS$ and $\gamma (v) \bigl(\x\bigr)$ $:=$
$\frac{\textstyle 1}{\textstyle 2} \, v \bigl(\x,\x\bigr)$
is a field from $\FT$.
\medskip

(c) If $v (\x,\y),$ $v'(\x,\y)$ $\in$ $\VS$ then
 there is a harmonic bilocal field
\beq\label{en4.7}
(v \PROD v') \bigl(\x,\y\bigr) := \,
\mathop{\text{w-\!}\lim}\limits_{\x' \, \to \, \y'} \
\bigl(\x'-\y'\bigr)^2
\Bigl( v\bigl(\x,\x'\bigr) \hspace{1pt} v'\bigl(\y',\y\bigr)
- \lvac v\bigl(\x,\x'\bigr) \hspace{1pt} v'\bigl(\y',\y\bigr) \rvac
\Bigr) .
\eeq
The existence of the above weak limit (i.e., a limit within correlation
functions)
will be established below together with the independence of $\x'=\y'$ and
the regularity of the resulting field for $(\x-\y)^2=0$.

\medskip

(d) If $v (\x,\y) \in \VS$ and $\phi (\x) \in \FT$ then we can construct
the following bilocal field belonging to $\VS$:
\beq\label{en4.7qq}
(v \PROD \phi) \bigl(\x,\y\bigr) := \,
\mathop{\text{w-\!}\lim}\limits_{\x' \, \to \, \y} \
\bigl(\x'-\y\bigr)^2
\Bigl( v\bigl(\x,\x'\bigr) \hspace{1pt} \phi \bigl(\y\bigr)
- \lvac v\bigl(\x,\x'\bigr) \hspace{1pt} \phi \bigl(\y\bigr) \rvac
\Bigr) ,
\eeq
where again the existence of the limit and the regularity for
$(\x-\y)^2=0$ will be established later.

One can define similarly a product $\phi \PROD v \in \VS$, but it would
then be expressed as: $(v^t \PROD \phi)^t$.

\medskip

To summarize, we have three bilinear maps:
$\FT \otimes \FT \mathop{\to}\limits^{\PROD} \VS$,
$\VS \otimes \VS \mathop{\to}\limits^{\PROD} \VS$,
$\VS \otimes \FT \mathop{\to}\limits^{\PROD} \VS$,
and two linear ones:
$\VS \mathop{\to}\limits^{t} \VS$, $\VS \mathop{\to}\limits^{\gamma} \FT$.
Applying these maps we construct $\FT$ and $\VS$ inductively,
starting from our original system of $d=2$ fields, given in
Theorem~\ref{T4.1}, and at each step of this inductive procedure,
we establish the existence of the above limits in (c) and (d).
In fact, we shall establish this together with the structure
of the truncated correlation functions for the fields in $\FT$ and
$\VS$.\footnote{%
Since we shall use the notion of truncated correlation functions also
for bilocal fields let us briefly recall it.
If $B_1,\dots,B_n$ are some smeared (multi)local fields then
their truncated correlation functions are recursively defined by:
$\lvac B_1 \cdots B_n \rvac$ $=$
$\mathop{\sum}\limits_{\dot{\cup} \, P \, = \, \{1,\dots,n\}}$
$\mathop{\prod}\limits_{\{j_1,\dots,j_k\} \, \in \, P}$
$\lvac B_{j_1} \cdots B_{j_k} \rvac\trn$
(the sum being over all partitions $P$ of $\{1,\dots,n\}$)}

Before we state the inductive result it is convenient to introduce
the vector space
\beq\label{AFE}
\AFE \, = \, \FT \times \VS
\eeq
and endow it with the following bilinear operation
\beq\label{MULT}
(\phi_1,v_1) \PROD (\phi_2,v_2) \, := \,
\bigl(0,\, \phi_1 \PROD \phi_2 + v_1 \PROD v_2 + v_1 \PROD \phi_2 + (v_2^t
\PROD \phi_1)^t \bigr),
\eeq
and with the transposition
\beq\label{INVOL}
(\phi,v)^t \, := \, (\phi,v^t) \,.
\eeq
The spaces $\FT$ and $\VS$ will be considered as subspaces in $\AFE$.
Thus, the new operation $\PROD$ in $\AFE$ combines
the above listed three operations.
We shall see later that $\AFE$ is actually an associative algebra
under the product~(\ref{MULT}).
We note that the transposition $t$ (\ref{INVOL}) is an \textit{antiinvolution}
with respect to the product:
$(q_1 \PROD q_2)^t = q_2^t \PROD q_1^t$,
for every $q_1,q_2 \in \AFE$.

\medskip

\noindent
\textbf{Proposition}~\thlabel{P4.3nn}
{\it
There exist multilinear functionals
\beq\label{C-funk}
c^{(N)} : \AFE^{\otimes N} \to \R
\eeq
such that if we take elements $q_1,\dots,q_{n+m} \in \AFE:$
$q_k$ $:=$ $v_k \bigl(\x_{k\hspace{0pt}[0]},$ $\x_{k\hspace{0pt}[1]}\bigr)
\in \VS$, where $[\varepsilon]$ stands for a $\Z/2\Z$--value and
$k = 1,\dots,n$, and
$q_k$ $:=$ $\phi_{k-n} \bigl(\x_k\bigr) \in \FT$
for $k = n+1,\dots,n+m$,
then the truncated correlation functions can be written in the following form:
\beqa\label{en4.10}
&& \hspace{-29pt}
\lvac
v_{1} \bigl(\x_{1\hspace{0pt}[0]},\x_{1\hspace{0pt}[1]}\bigr) \cdots
v_{n} \bigl(\x_{n\hspace{0pt}[0]},\x_{n\hspace{0pt}[1]}\bigr)
\,
\phi_1 \bigl(\x_{n+1}\bigr) \cdots \phi_{m} \bigl(\x_{n+m}\bigr)
\rvac\trn
\nn && \hspace{-29pt} \hspace{2pt} =
\frac{1}{2(n \hspace{-1pt}+ \hspace{-1pt} m)}
\hspace{-5pt}
\mathop{\sum} \limits_{\mathop{}\limits^{\sigma \, \in \,
\mathcal{S}_{n+m}}_{(\varepsilon_1,\dots,\varepsilon_n) \, \in \, (\Z/2\Z)^n}}
\hspace{-7pt}
K_{\sigma,\varepsilon} \ T_{\sigma,\varepsilon}
\bigl(\x_{1[0]},\dots,\x_{n[1]},\x_{n+1},\dots,\x_{n+m}\bigr)^{-1}\!\! .
\quad
\eeqa
Here: $K_{\sigma,\varepsilon}$ are coefficients given by
$K_{\sigma,\varepsilon} := c^{(n+m)}
\Bigl(q_{\sigma_1}^{[\varepsilon_{\sigma_1}]},
\dots,q_{\sigma_{n+m}}^{[\varepsilon_{\sigma_{n+m}}]}\Bigr)$,
where we set $\varepsilon_{n+1}=\cdots=\varepsilon_{n+m}=0$, and
$q^{[0]} := q,$ $q^{[1]} := q^t$ (for $q \in \AFE$); the terms
$T_{\sigma,\varepsilon}$ are the following cyclic products of intervals
\beqa\label{T-term}
T_{\sigma,\varepsilon} \, &=&
\bigl(\x_{\sigma_{n+m}}-\x_{\sigma_{1}\hspace{0pt}[\varepsilon_1]} \bigr)^2 \
\mathop{\prod}\limits_{k \, = \, 1}^{n-1}
\bigl(\x_{\sigma_{k}\hspace{0pt}[1+\varepsilon_k]}
- \x_{\sigma_{k+1}\hspace{0pt}[\varepsilon_{k+1}]} \bigr)^2
\nn && \times \,
\bigl(\x_{\sigma_{n}\hspace{0pt}[1+\varepsilon_n]}-\x_{\sigma_{n+1}} \bigr)^2
 \ \mathop{\prod}\limits_{k \, = \, 1}^{m-1}
\bigl(\x_{\sigma_{n+k}} - \x_{\sigma_{n+k+1}} \bigr)^2 \,.
\eeqa
It follows by Eq.~(\ref{en4.10}) that the limits in the steps
(c) and (d) above are well defined.}

\medskip

Before the proof let us make some remarks.
First, we used the same notation $c^{(n)}$ as in Theorem~\ref{T4.2}
since the above multilinear functionals are obviously an extension of the
previous, i.e., Eq.~(\ref{en4.10}) reduces to Eq.~(\ref{eq5.1b}) for $m=0$.
Let us also give an example for Eq.~(\ref{en4.10}) with $n=m=1$:
\beqa\label{EXA}
&
\lvac v(\x_1,\x_2) \phi (\x_3) \rvac \, = \,
\frac{\textstyle 1}{\textstyle 4}
\, \Bigl( c^{(2)} (v,\phi)\,\bigl(\rho_{23}\,\rho_{31}\bigr)^{-1}
+ c^{(2)} (v^t,\phi)\,\bigl(\rho_{13}\,\rho_{32}\bigr)^{-1}
\hspace{-10pt} & \nn &
+ \, c^{(2)} (\phi,v)\,\bigl(\rho_{31}\,\rho_{23}\bigr)^{-1}
+ c^{(2)} (\phi,v^t)\,\bigl(\rho_{32}\,\rho_{13}\bigr)^{-1}
\Bigr) .
\hspace{-10pt} &
\eeqa

As one can see, $c^{(n)}$ (as well as $c^{(n)}$ of Theorem~\ref{T4.2})
possess a cyclic and an inversion symmetry:
\beq\label{en4.23nn}
c^{(n)} \bigl(q_1,\dots,q_n\bigr) \, = \,
c^{(n)} \bigl(q_n,q_1\dots,q_{n-1}\bigr) \, = \,
c^{(n)} \bigl(q_n^t,\dots,q_1^t\bigr) \,.
\eeq
This is the reason for choosing the prefactors in Eqs.~(\ref{eq5.1b})
and~(\ref{en4.10}) (the inverse of the orders of the symmetry groups).

\medskip

\noindent
\textit{Proof of Proposition~\ref{P4.3nn}.}
According to our preliminary remarks it is enough to prove that
Eq.~(\ref{en4.10}) is consistent with the operations
$\FT \otimes \FT \mathop{\to}\limits^{\PROD} \VS$,
$\VS \otimes \VS \mathop{\to}\limits^{\PROD} \VS$,
$\VS \otimes \FT \mathop{\to}\limits^{\PROD} \VS$ and
$\VS \mathop{\to}\limits^{\gamma} \FT$.

Starting with $\FT \otimes \FT \mathop{\to}\limits^{\PROD} \VS$ one
should prove that any truncated correlation function
$\La \cdot \phi_1 (\x_1) \, \phi_2 (\x_2) \cdot \Ra\trn$
given by Eq.~(\ref{en4.10}) yields a harmonic decomposition:
$\rho_{12} \, \La \cdot \phi_1(\x_1) \phi_2(\x_2) \cdot \Ra\trn
= \La \cdot (\phi_1 \PROD \phi_2)\bigl(\x_1,\x_2\bigr) \cdot \Ra\trn +
\rho_{12} \, R(\x_1,\x_2)$, with a correlation function
$\La \cdot (\phi_1 \PROD \phi_2)\bigl(\x_1,\x_2\bigr) \cdot \Ra\trn$ given by
Eq.~(\ref{en4.10}) and a rational function $R$ regular at $\rho_{12} = 0$.
This gives us relations of the type
\beq\label{en4.17nn1}
c^{(n+2)} (q_1,\dots,\phi_1,\phi_2,\dots,q_n) \, = \,
c^{(n+1)} (q_1,\dots,\phi_1\PROD\phi_2,\dots,q_n) \,.
\eeq

Next, having correlation functions of type
$\La \cdot v_1(\x_1,\x_2)  v_2(\x_3,\x_4) \cdot \Ra\trn$ or
$\La \cdot v(\x_1,\x_2) \phi(\x_3) \cdot \Ra\trn$ of the form
(\ref{en4.10}), one verifies that the limits (\ref{en4.7}) and
(\ref{en4.7qq}) exist within these correlation functions, and they
yield expressions for $\La \cdot (v_1 \PROD v_2) \bigl(\x_1,\x_4\bigr)
\cdot \Ra\trn$ and $\La \cdot (v \PROD \phi) \bigl(\x_1,\x_3\bigr)
\cdot \Ra\trn$ consistent with (\ref{en4.10}). As a result we obtain
again relations between the $c$'s: 
\beqa\label{en4.18nn1}
c^{(n+2)} (q_1,\dots,v_1,v_2,\dots,q_n) \, &=&
c^{(n+1)} (q_1,\dots,v_1\PROD v_2,\dots,q_n) \,,
\nn
c^{(n+2)} (q_1,\dots,v,\phi,\dots,q_n) \, &=&
c^{(n+1)} (q_1,\dots,v\PROD \phi,\dots,q_n) \,.
\eeqa

Finally, one verifies that setting $\x_1=\x_2$ in $\La \cdot
v(\x_1,\x_2) \cdot \Ra\trn$ we obtain the correlation functions $\La
\cdot \gamma(v)\bigl(\x_1\bigr) \cdot \Ra\trn$ with the relation
\beq\label{en4.19nn1}
c^{(n+1)} (q_1,\dots,(v+v^t),\dots,q_n) \, = \,
2\;c^{(n+1)} (q_1,\dots,\gamma(v),\dots,q_n) \,.
\eeq
This completes the proof of Proposition~\ref{P4.3nn} as well as
the proof that the products $\VS \otimes \VS \mathop{\to}\limits^{\PROD} \VS$
and $\VS \otimes \FT \mathop{\to}\limits^{\PROD} \VS$ are well defined.
\hfill$\Box$

\subsection{Associative algebra structure of the OPE}
\label{Sse4.2nn}

Note that Eqs.~(\ref{en4.17nn1}), (\ref{en4.18nn1}) read (under (\ref{MULT}))
\beq\label{en4.24nn}
c^{(n)} \bigl(q_1,\dots,q_k, q_{k+1},\dots,q_n\bigr) \, = \,
c^{(n-1)} \bigl(q_1,\dots,q_k \PROD q_{k+1},\dots,q_n\bigr) \,.
\eeq
This implies that
\textit{the bilinear operation $\PROD$ on $\AFE$ is an associative product}.

Indeed, consider the element $q:=\bigl(q_1 \PROD q_2\bigr) \PROD q_3-
q_1 \PROD \bigl(q_2 \PROD q_3\bigr)$ for $q_1,q_2,q_3 \in \AFE$.
By (\ref{MULT}) $q$ is a bilocal field.
Equation (\ref{en4.24nn}) implies that all $c$'s in which $q$ enters
vanish and hence, by Eq.~(\ref{en4.10}) $q$ has zero correlation functions
with all other fields, including itself.
But then this (bilocal) field is zero by the Reeh--Schlieder theorem,
since its action on the vacuum will be identically zero.

Thus, introducing the cartesian product $\AFE$~(\ref{AFE}) was not
only convenient for combining three types of bilinear operations in one
but also as a compact expression for the associativity
(Eqs.~(\ref{en4.17nn1}), (\ref{en4.18nn1})). However, $\AFE$ carries a
redundant information due to the following relation:
\beq\label{REL}
\bigl(-\gamma(v),\,\frac{1}{2} \hspace{1pt} (v+v^t)\bigr)*q \,=\, 0 \,=\,
q * \bigl(-\gamma (v),\,\frac{1}{2} \hspace{1pt} (v+v^t)\bigr) \,
\eeq
for every $v \in \VS$ and $q \in \AFE$.
To prove (\ref{REL}) we point out first that it is equivalent to the
identities $v \PROD \phi$ $=$ $\gamma (v) \PROD \phi$
and $v' \PROD v$ $=$ $v' \PROD \gamma (v)$
for $v$ $=$ $v^t \in \VS$ and any $\phi \in \FT$, $v' \in \VS$.
These identities can be established again first for the $c$'s,
and then proceeding by using the Reeh--Schlieder theorem,
as in the above proof of associativity.

Hence, the redundancy in $\AFE$ is because we can identify symmetric
bilocal fields $v = v^t \in \VS$ with their restrictions to the diagonal,
$\gamma (v) \in \FT$, and this is compatible with the product $\PROD$.
Let us point out that the restriction of the map $\gamma$ to the
$t$--invariant subspace
$\VS_s$ $:=$ $\{v \in \VS$ $:$ $v=v^t\}$ is an injection into $\FT$.
The latter follows from a simple analysis
of the $4$-point functions of $v$ and the Reeh--Schlieder theorem:
if $v (\x,\y)$ $=$ $v (\y,\x)$ and $\lvac v (\x,\x) v (\y,\y) \rvac$ $=$ $0$
then $\lvac v (\x,\x') v (\y,\y') \rvac$ $=$ $0$.
In this way we see that we can identify in $\AFE$ the symmetric
harmonic bilocal fields
$v=v^t$ with their restriction on the diagonal $\gamma (v) \in \FT$.

Formally, the above considerations can be summarized in the following
abstract way. Let us introduce the quotient
\beq\label{AF}
\AF \, := \, \AFE \Bigl/ \Bigl\{\bigl(-\gamma (v),\,\frac{1}{2}
\hspace{1pt} (v+v^t)\bigr): v \in \VS\Bigr\} \,.
\eeq
It is an associative algebra according to Eq.~(\ref{REL}).
The involution $t : \AFE \to \AFE$ can be transferred to an involution
on the quotient (\ref{AF}) and we denote it by $t$ as well.
The spaces $\FT$ and $\VS$ are mapped into $\AF$ by
the natural compositions $\FT \to \AFE \to \AF$ and $\VS \to \AFE \to \AF$.
The injectivity of $\gamma$ on $\VS_s$ implies that the maps
$\FT \to \AF$ and $\VS \to \AF$ so defined are actually \textit{injections}.
Hence, we shall treat $\FT$ and $\VS$ also as subspaces of $\AF$.
Furthermore, $\AF$ becomes a direct sum of vector spaces
\beqa\label{en4.25nn11}
\AF \, &=& \FT \oplus \VS_a
\, , \quad
\\
\label{en4.26nn11}
\text{with} \quad \qquad
\bigl\{q \in \AF : q^t \, = \, q\bigr\} \, &=& 
\FT \, \supseteq \,
\VS_s \ \big( \, := \bigl\{v \in \VS : v^t = v\bigr\} \big)
, \quad
\nn
\bigl\{q \in \AF : q^t \, = \, - q\bigr\} \, &=&
\VS_a \, := \,
\bigl\{v \in \VS : v^t = - v\bigr\}.
\nonumber
\eeqa
Hence, the $t$--symmetric elements of $\AF$ are identified with the
$d=2$ local fields, while the $t$--antisymmetric elements of $\AF$,
with the antisymmetric, harmonic bilocal $(1,1)$ fields.
(Neither $\FT$ nor $\VS_a$ are sub{\em algebras} of $\AF$.)

To summarize, the associative algebra $\AF$ is obtained from $\AFE$
by identifying the space $\VS_s$ of symmetric bilocal fields with its image
$\gamma \bigl(\VS_s\bigr) \subseteq \FT$.

For simplicity we will denote the equivalence class in
$\AF$ of an element $q \in \AFE$ again by $q$.
Also note that the $c$'s can be transferred as well, to multilinear
functionals on $\AF$, since the kernel of the quotient~(\ref{AF})
is contained in the kernel of each $c^{(n)}$ by (\ref{en4.19nn1}).
We shall use the same notation $c^{(n)}$ also for
the multilinear functional $c^{(n)}$ on $\AF$.

\medskip

\noindent
\textbf{Example~4.1.}
Let us illustrate the above algebraic structures on the simplest example of
a QFT generated by a pair of $d=2$ GCI fields $\Phi_1$ and $\Phi_2$
given by normal a pair of two mutually commuting free massless
fields $\varphi_j$:
$\Phi_1 (\x)$ $=$ $\frac{\textstyle 1}{\textstyle 2} \,
\bigl(:\!\varphi_1^2 (\x)\!:-:\!\varphi_2^2 (\x)\!:\bigr)$ and
$\Phi_2 (\x)$ $=$ $\varphi_1 (\x) \, \varphi_2 (\x)$.
Their OPE algebra involves a set of four independent harmonic bilocal fields
$V_{jk} (\x_1,\x_2)$ $:=$ $:\!\varphi_j (\x_1) \, \varphi_k (\x_2)\!:$
($j,k=1,2$), which satisfy
$\bigl[V_{jk}(\x_1,\x_2)\bigr]^*$ $=$
$V_{kj}(\x_1,\x_2)$ $=$ $V_{jk}(\x_2,\x_1)$.
For instance, we have $\Phi_1 \PROD \Phi_2$ $=$
$V_{12} - V_{21}$.\footnote{%
i.e., in the OPE $\Phi_1 (\x_1) \Phi_2 (\x_2)$ there appears the
antisymmetric bilocal field
$V_{12} (\x_1,\x_2)$ $-$ $V_{21} (\x_1,\x_2)$ that involves only odd rank
conserved tensor currents in its expansion in local fields}
Also note that $\Phi_1$ $=$ $\gamma (V_1)$ for $V_1(\x_1,\x_2)$ $=$
$:\!\varphi_1(\x_1) \, \varphi_1(\x_2)\!:-
:\!\varphi_2(\x_1) \, \varphi_2(\x_2)\!:$, etc.

\medskip

By the associativity and Eq.~(\ref{en4.24nn}) we have
\beq\label{en4.25nn}
c^{(n)} \bigl(q_1,\dots,q_n\bigr) \, = \,
c^{(2)} \bigl(q_1 \PROD \cdots \PROD q_{n-1} , q_n\bigr)
\eeq
for $q_1,\dots,q_n \in \AF$.
Let us consider now $c^{(2)}$ and define the following \textit{symmetric}
bilinear form on $\AF$:
\beq\label{BiFo}
\La q_1,q_2 \Ra \, := \, c^{(2)} \bigl(q_1^t,q_2\bigr) \,.
\eeq
First note that $\FT$ and $\VS_a$ are orthogonal with respect to
this bilinear form:
this is due to the fact that there is no nonzero
three point conformally invariant scalar function
of weights $(2,1,1)$, which is antisymmetric in the second and third
arguments. Next, we claim that (\ref{BiFo}) is strictly positive definite.
This is a straightforward consequence of the Wightman positivity and the
Reeh--Schlieder theorem
(one should consider separately the positivity on $\FT$ and $\VS_a$).
In particular, (\ref{BiFo}) is nondegenerate.
By Eqs.~(\ref{en4.23nn}) and (\ref{en4.24nn}) we have:
\beq\label{en4.28nn3}
\La q_1 \PROD q_2,\, q_3 \Ra \, = \, \La q_2,\, q_1^t \PROD q_3 \Ra
\eeq
for all $q_1,q_2,q_3 \in \AF$.

Let us introduce now an additional splitting of $\FT$.
Denote by $\FT_0$ the kernel of the product, i.e.,
\beq\label{en4.28nn1}
\FT_0 := \bigl\{ \psi \in \FT : \psi \PROD q = 0 \; \forall q \in \AF \bigr\}
\equiv \bigl\{ \psi \in \FT : q \PROD \psi = 0 \; \forall q \in \AF \bigr\}
\eeq
(the second equality is due to the identity
$\phi \PROD q$ $=$ $(q^t \PROD \phi)^t$).
Let $\FT_1$ be the orthogonal complement in $\FT$
of $\FT_0$ with respect to the scalar product (\ref{BiFo}):
\beq\label{en4.29nn1}
\FT_1 := \bigl\{ \phi \in \FT : \La \phi, \psi \Ra = 0 \;
\forall \psi \in \FT_0 \bigr\} \,.
\eeq

The meaning of fields belonging to $\FT_0$ becomes immediately clear if
we note that $c^{(n)}$ for $n \geqslant 3$ are zero if one of the
arguments belongs to $\FT_0$ (this is due to Eq.~(\ref{en4.25nn})).
Hence, all their truncated functions higher than two point are zero, i.e.,
the fields belonging to $\FT_0$ are \textit{generalized free $d=2$ fields}.
Furthermore, these fields commute with all other fields from $\FT_1$
and $\VS_a$ $\equiv$ $\AF^{(1)}$:
this is because of the vanishing of $c^{(2)} (\psi,q)$ if $\psi \in \FT_0$ 
and $q \in \FT_1 \oplus \VS_a$, 
as well as of all $c^{(n+1)} (\psi,q_1,\dots,q_n)$
for $n\geq 2$ if $\psi \in \FT_0$ and $q_1,\dots,q_n \in \AF$
(by (\ref{en4.25nn}) and (\ref{en4.28nn1})).

Clearly, $\FT_1 \oplus \VS_a$ is a subalgebra of $\AF$:
this follows from Eq.~(\ref{en4.28nn3}) with $q_3\in\FT_0$ along with
the definitions (\ref{en4.28nn1}) and (\ref{en4.29nn1}).
Let us denote it by
\beq\label{QQ}
\QQ \, := \, \FT_1 \oplus \VS_a \,.
\eeq
We are now ready to state the main step towards the proof of
Theorem~\ref{T4.1}.

\medskip

\noindent
\textbf{Proposition}~\thlabel{P4.5nn1}
{\it
There is a homomorphism $\iota$ from the associative algebra $\QQ$
into the algebra of Hilbert--Schmidt operators over some real
separable Hilbert space, such that
\beq\label{en4.30nn1}
c^{(n)} \bigl(q_1,\dots,q_n\bigr) \, = \, \text{\rm Tr} \,
\Bigl( \iota \bigl(q_1\bigr) \cdots \iota \bigl(q_n\bigr) \Bigr)
\,,
\eeq
and $\iota \bigl(\FT\bigr)$ are symmetric operators while
$\iota \bigl(\VS_a\bigr)$ are antisymmetric.}

\medskip

We shall give the proof of this proposition in the subsequent subsection.
The main reason leading to it is that $\QQ$ becomes a real
\textit{Hilbert algebra} with an \textit{integral} trace on it.
Here we proceed to show how Theorem~\ref{T4.1} can be proven by
using the above results.

\medskip

\noindent
\textit{Proof of Theorem~\ref{T4.1}.}
Let $\Phi_m$ $=$ $\Phi_m^0$ $+$ $\Phi_m^1$ be the decomposition of each
field $\Phi_m$ according to the splitting $\FT = \FT_0 \oplus \FT_1$.
Take an orthonormal basis $\psi_m$ in $\FT_0$ and let
$\Phi_m^0$ $=$
$\mathop{\sum}\limits_{j \,=\, 1}^{\infty}$ $\alpha_{m,j}$ $\psi_j$,
and $\beta_m$ $=$ $\bigl(\beta_{m,j,k}\bigr)_{j,k}$ be the symmetric matrix
corresponding to the Hilbert--Schmidt operator $\iota \bigl(\Phi_m^1\bigr)$
($m=1,2,\dots$).
Then Eqs.~(\ref{eq5.3nn}) and (\ref{en4.30nn1})
show that the constants $\alpha_{m,j}$ and $\beta_{m,j,k}$ so defined
satisfy the conditions of Theorem~\ref{T4.1}.~\hfill$\Box$

\medskip

\noindent
\textit{Remark~4.1.}
In general, we have $\FT_1 \supsetneqq \VS_s$.
This is because the elements of $\FT_1$ correspond,
by Proposition~\ref{P4.5nn1}, to Hilbert--Schmidt symmetric operators
and on the other hand, the elements of $\VS$
are obtained, according to the inductive construction of Sect.~\ref{Sse4.1},
as products of elements of $\FT$
and will, hence, correspond to trace class operators.

\subsection{Completion of the proofs}\label{Sse4.3nn}

It remains to prove Proposition~\ref{P4.5nn1}.
We start with an inequality of Cauchy--Schwartz type.

\medskip

\noindent
\textbf{Lemma}~\thlabel{L4.5n1}
{\it
Let $q_1,q_2 \in \AF$ be such that each of them belongs either to 
$\FT$ or to $\VS_a$. Then we have
\beq\label{xx4.28}
\La q_1 \PROD q_2 , q_1 \PROD q_2 \Ra^2 \leqslant
\La q_1 \PROD q_1 , q_1 \PROD q_1 \Ra \,
\La q_2 \PROD q_2 , q_2 \PROD q_2 \Ra \,.
\eeq}

\noindent
\textit{Proof.}
Consider
\(
\La q_1 \PROD q_1 + \lambda \, q_2 \PROD q_2,
q_1 \PROD q_1 + \lambda \, q_2 \PROD q_2 \Ra \geqslant 0
\)
and use that
$\La q_1 \PROD q_1 , q_2 \PROD q_2 \Ra$ $=$
$\pm$ $\La q_1 \PROD q_2 , q_1 \PROD q_2 \Ra$
if each of  $q_1,q_2$ belongs either to $\FT$ or to $\VS_a$.~\hfill$\quad\Box$

\medskip

The space $\QQ$~(\ref{QQ}) is a real pre--Hilbert space with a scalar product
given by (\ref{BiFo}).
It is also invariant under the action of $t$
(actually the eigenspaces of $t$ are $\FT_1$ and $\VS_a$).
The left action of $\QQ$ on itself gives us an algebra homomorphism
\beq\label{en4.33nn2}
\iota : \QQ \to \text{\rm Lin}_{\R} \, \QQ
\eeq
of $\QQ$ into the algebra of all operators over $\QQ$.
Moreover, the elements of $\FT$ are mapped into symmetric operators
and the elements of $\VS_a$, into antisymmetric
(this is due to (\ref{en4.28nn3})).

\medskip

\noindent
\textbf{Lemma}~\thlabel{L4.6nn}
{\it
Every element of $\QQ$ is mapped into a Hilbert--Schmidt operator.}

\medskip

\noindent
\textit{Proof.} Since $\QQ$ is generated by $\FT_1$
(according to the inductive construction of $\FT$ and $\VS$ in
Sect.~\ref{Sse4.1}) it is enough to show this for the elements of  $\FT_1$.

Let $\phi \in \FT_1$ and consider the commutative subalgebra $\QQ_{\phi}$
of $\QQ$ generated by $\phi$.
The algebra $\QQ_{\phi}$ is freely generated by $\phi$,
i.e., is isomorphic to the algebra $\lambda \, \R[\lambda]$
of polynomials in a single variable $\lambda$ ($\leftrightarrow \phi$),
since $\phi$ belongs to the orthogonal complement of $\FT_0$~(\ref{en4.28nn1}).
For a $p (\lambda) \in \lambda \, \R [\lambda]$ we shall denote by
$\phi^{[p]}$ the corresponding element of $\QQ_{\phi}$.
In particular,
\beq\label{en4.32nn1}
\phi^{[p_1]} \PROD \phi^{[p_2]} \, = \, \phi^{[p_1 p_2]}
.
\eeq
Setting
\beq\label{en4.32nn}
\phi^{\PROD (n+1)} := \phi^{\PROD n} \PROD \phi , \quad
c \bigl[\lambda^{n+1}\bigr] := c^{(2)} \bigl(\phi^{\PROD n},\phi\bigr)
\equiv \La \phi^{\PROD n},\phi \Ra
\eeq
($\phi^{\PROD 1} := \phi$, $n \geqslant 1$) we obtain a \textit{positive}
definite functional over the algebra
$\lambda^2 \, \R [\lambda]$ $\cong$ $\phi \PROD \QQ_{\phi}$
(due to Eq.~(\ref{en4.28nn3}) and the positivity of
$\La \cdot,\cdot \Ra$ (\ref{BiFo})).

Then, by the Hamburger theorem about the classical moment problem
(\cite[Chap.~12, Sect.~8]{DS63})
we conclude that there exists a bounded positive Borel measure
$\mathrm{d}\mu \bigl( \lambda \bigr)$ on $\R$, such that
\beq\label{eq5.15d}
c \left[ \, \lambda^2 \, p \left( \lambda \right) \right] \, = \,
\mathop{\int}\limits_{\!\!\!\!\!\! \R}
p \bigl(\lambda\bigr) \,
d\mu(\lambda)  \
\eeq
for every $p(\lambda) \in \R[\lambda]$.
Using this we can extend the fields $\phi^{[p]} (\x)$
to $\phi^{[f]} (\x)$ for Borel measurable functions
$f$ having compact support with respect to $\mu$ in $\R \backslash \{0\}$.
The latter can be done in the following way.
Fix $\varepsilon \in (0,1)$ and let $g_1,\dots,g_n$ be Schwartz test
functions on $M$. By Theorem~\ref{T4.2} the correlators
$\lvac \phi^{[p_1]} [g_1] \cdots \phi^{[p_n]} [g_n] \rvac$ depend
polynomially on
$c^{(n)} \bigl(\phi^{[p_{k_1}]},\dots,\phi^{[p_{k_j}]}\bigr)$ $=$
$c \bigl[p_{k_1}\cdots p_{k_j}\bigr]$ for all
$\{k_1,\dots,k_j\} \subseteq \{1,\dots,n\}$.
But for every $\varepsilon \in \left( 0,\, 1 \right)$ there exists a norm
\beq\label{q-norm}
\| q \|_{\varepsilon} \, = \, A_{\varepsilon} \
\mathop{\sup}\limits_{|\lambda| \, \leq \, \varepsilon}
\ \Bigl| \frac{q_k(\lambda)}{\lambda^2} \Bigr|
\, + \, B_{\varepsilon} \ \
\mathop{\int}\limits_{\!\!\!\!\!\!
\R \, \backslash \, \left( -\varepsilon,\, \varepsilon \right)}
\bigl| q_k(\lambda) \bigr| \, d\mu(\lambda)
\eeq
on $\lambda^2 \R [\lambda] \ni q(\lambda)$,
where $A_{\varepsilon}$ and $B_{\varepsilon}$ are some positive constants,
such that for every $q_1,\dots,q_m \in \lambda^2\R[\lambda]$
\beq\label{+?}
\Bigl| \, c \bigl[q_1(\lambda) \cdots q_m (\lambda) \bigr] \Bigr|
\leq
\mathop{\prod}\limits_{k \, = \, 1}^m \,
\Bigl\{
\mathop{\int}\limits_{\!\!\!\!\!\! \R} \,
\frac{\bigl| q_k(\lambda) \bigr|^m}{|\lambda|^2} \,
d\mu(\lambda)
\Bigr\}^{\frac{1}{m}}
\leqslant \,
\mathop{\prod}\limits_{k \, = \, 1}^m \
\| q_k \|_{\varepsilon}
\, .
\nonumber
\eeq
\vskip-2mm\noindent
Hence,
$\bigl|\lvac \phi^{[p_1]} [g_1] \cdots \phi^{[p_n]} [g_n] \rvac\bigr|
\, \leqslant \,
C \, \mathop{\prod}\limits_{k \, = \, 1}^n \
\| p_k \|_{\varepsilon} \, \|g_k\|_S $
for some constant $C$ and Schwartz norm $\|$$\cdot$$\|_S$ (not
depending on $p_k$ and $g_k$).
Since for every $\varepsilon \in (0,1)$ the Banach space
$L^1 \bigl(\R \backslash \{(-\varepsilon,\varepsilon)\},\mu\bigr)$
is contained in the completion of $\lambda^2 \R[\lambda]$ with respect to the
norms (\ref{q-norm}), we can extend the linear functional
$c [p(\lambda)]$ as well as the correlators
$\lvac \phi^{[p_1]} [g_1] \cdots \phi^{[p_n]} [g_n] \rvac$
to a functional $c [f(\lambda)]$ and correlators
$\lvac \phi^{[f_1]} [g_1] \cdots \phi^{[f_n]} [g_n] \rvac$ defined for
Borel functions $f,f_1,\dots,f_n$ compactly supported with respect to
$\mu$ in $\R \backslash \{0\}$.
Thus, we can extend the fields $\phi^{[p]}$ by extending their correlators.

By the continuity we also have for arbitrary Borel functions $f,f_k$,
compactly supported in $\R\backslash \{0\}$:
\beqa\label{eq5.22h}
\phi^{[f_1]} \PROD \phi^{[f_2]} = \phi^{[f_1f_2]}
, \quad &&
c^{(n)} \bigl( \phi^{[f_1]} , \dots, \phi^{[f_n]} \bigr)
= c\bigl[f_1 \cdots f_n\bigr],\nn &&
c\bigl[f] = \mathop{\int}\limits_{\!\!\!\!\!\! \R}
\frac{f (\lambda)}{\lambda^2} \ d \mu (\lambda)
\eeqa
\vskip-2mm\noindent
(cp. (\ref{eq5.15d})),
and $c^{(n)}$ determine the correlation functions of $\phi^{[f_k]}$
as in Theorem~\ref{T4.2}.

In particular, for every characteristic function $\chi_{S}$ of a compact subset
$S \subset \R \backslash \{0\}$ we have $\phi^{[\chi_{S}]}
\PROD \phi^{[\chi_{S}]} = \phi^{[\chi_{S}]}$.
Hence, for such a $d=2$ field we will have that all its truncated
correlation functions are given by (\ref{eq5.1b}) with all normalization
constants $c^{(n)}$ equal to one and the same value
$c^{(2)} \bigl(\phi^{[\chi_{s}]},\phi^{[\chi_{s}]}\bigr)$.
Then, as shown in \cite[Theorem 5.1]{NST02}, Wightman positivity
requires this value to be a non-negative integer, i.e.,
\beq\label{eqYY}
c^{(2)} \bigl( \phi^{[\chi_{S}]} , \phi^{[\chi_{S}]} \bigr) \, = \,
c \bigl[\chi_{S}\bigr] \, = \,
\mathop{\int}\limits_{\!\!\!\!\!\! S} \frac{d\mu(\lambda)}{\lambda^2} \,
\in \, \{0,1,2,\dots\}
\eeq
\vskip-2mm\noindent
(it is zero iff $\phi^{[\chi_{S}]} = 0$).
Hence, the restriction of the measure $d\mu (\lambda)/\lambda^2$ to
$\R \backslash\{0\}$ is a (possibly infinite) sum of
atom measures of integral masses, each supported at
some $\gamma_k \in \R\backslash\{0\}$ for $k=1,\dots,N$ (and $N$ could
be infinity).
In particular, the measure $\mu$ is supported in a bounded subset of $\R$.

By Lemma~\ref{L4.5n1} we can define $\iota (\phi^{[f]})$
as a closable operator on $\QQ$ if $f$ is a Borel measurable function 
with compact support in $\R \backslash \{0\}$.
It follows then that the projectors $\iota (\phi^{[\chi_{S}]})$,
for a compact $S \subseteq \R \backslash \{0\}$,
provide a spectral decomposition for $\iota (\phi)$
(in fact, $\iota (\phi^{[f]})$ $=$ $f \bigl(\iota (\phi)\bigr)$).
Thus, $\iota (\phi)$ has discrete spectrum with
eigenvalues
$\gamma_k$ ($k \in \N$),
each of a
multiplicity given by the integer
$c^{(2)}\bigl(\phi^{\chi_{\{\gamma_k\}}},\phi^{\chi_{\{\gamma_k\}}}\bigr)$.
Then $\iota (\phi)$ is a Hilbert--Schmidt operator since
\beqa\label{HiSc}
\mathop{\sum}\limits_{k \, = \, 1}^{\infty} \,
\gamma_k^2 \ c^{(2)} \bigl( \phi^{\chi_{\{\gamma_k\}}} ,
\phi^{\chi_{\{\gamma_k\}}} \bigr) \, =
\mathop{\sum}\limits_{k \, = \, 1}^{\infty} \,
\gamma_k^2 \,
\mathop{\int}\limits_{\!\!\!\! \{\gamma_k\}} \frac{d\mu(\lambda)}{\lambda^2}
\, =
\mathop{\int}\limits_{\! \R \backslash \{0\}} d \mu (\lambda)
< \infty
\nonumber
\eeqa
($\mu$ being a bounded measure).~\hfill$\Box$

\medskip

The completion of the proof of Proposition~\ref{P4.5nn1}
is provided now by the following corollary.
\medskip

\noindent
\textbf{Corollary}~\thlabel{C4.7nn}
{\it For every
$q_1,q_2 \in \QQ$ one has
$c^{(2)} \bigl(q_1,q_2\bigr)$ $=$ $\text{\rm Tr} \hspace{1.5pt}
\bigl(\iota (q_1) \iota (q_2)\bigr)$.}

\medskip

\noindent
\textit{Proof.}
If $q_1=q_2 \in \FT_1$ this follows from the proof of Lemma~\ref{L4.6nn}
and hence, by a polarization, for any $q_1,q_2 \in \FT_1$.
The general case can be obtained by using the facts
that $\QQ$ is generated by $\FT_1$ and
$c^{(2)}$ has the symmetry $c^{(2)} (q_1 \PROD q_2, q_3)$ $=$
$c^{(2)} (q_1, q_2 \PROD  q_3)$.~\hfill$\Box$

\section{Discussion. Open problems}\label{Sse5nn}

The main result of Sect.~\ref{Se5},
the (generalized) free field representation of a system $\{\phi_a\}$
of GCI scalar fields of conformal dimension $d=2$ (Theorem~\ref{T4.1}),
is obtained by revealing and exploiting a rich algebraic structure
in the space $\FT \times \VS$ of all $d=2$ real scalar fields and
of all harmonic bilocal fields of dimension $(1,1)$.
However, this structure is mainly due to the fact that
we are in the case of lower scaling dimension:
there is only one possible singular structure in the OPE
(after truncating the vacuum part).
One can try to establish such a result in spaces of \textit{spin--tensor}
bilocal fields
(of dimension $\bigl(\frac{3}{2},\frac{3}{2}\bigr)$ or $\bigl(2,2\bigr)$)
satisfying linear (first order) conformally invariant differential equations
(that again imply harmonicity).
If these equations together with the corresponding pole bounds
imply such singularities in the OPE, which can be ``split''
one would be able to prove the validity of free field realizations
in such more general theories, too.

One may also attempt to study models,
say in a theory of a system of scalar fields of dimension $d=4$,
without leaving the realm of scalar bilocal harmonic fields $V_1$
(of dimension $(1,1)$).
In \cite{NRT05} there have been found examples of $6$--point functions
of harmonic bilocal fields, which do not have free field realizations.
However, our experience with the $d=2$ case shows that in order to
complete the model
(including the check of Wightman positivity for all correlation functions)
it is crucial to describe the OPE in terms of some simple algebraic structure
(e.g., associative, or Lie algebras).

On the other hand going beyond bilocal $V_1$'s is a true signal of
\textit{nontriviality} of a GCI model.
Our analysis of Sect.~\ref{Se4} shows that this can be characterized
by a simple property of the correlation functions: the violation of the
single pole property (of Sect.~\ref{Sse3.3n}).
From this point of view a further exploration of the example
of Sect.~\ref{Sse3.4} within a QFT involving currents appears
particularly attractive.

\bigskip

\noindent
{\textbf{Note added in proof.} In \cite{NRT07}, we have determined the
  biharmonic function whose leading part is given by Eq.\ (\ref{equ3.30x}). It
  involves dilogarothmic functions, whose arguments are algebraic
  functions of conformal cross ratios. This exemplifies the violation
  of Huygens bilocality for the biharmonic fields, Theorem 3.7. Yet,
  in support of Conjecture 3.3, it is shown that the structure of the
  cuts is in a nontrivial manner consistent with ordinary bilocality. 

\bigskip

\noindent
{\textbf{Acknowledgements.}

We thank Yassen Stanev for an enlightening discussion.

This work was started while N.N. and I.T. were visiting the
Institut f\"ur Theoretische Physik der Universit\"at G\"ottingen as an
Alexander von Humboldt research fellow and an AvH awardee, respectively.
It was continued during the stay of N.N.
at the Albert Einstein Institute for Gravitational Physics in Potsdam
and of I.T. at the Theory Group of the Physics Department of CERN.
The paper was completed during the visit of N.N. and I.T. to
the High Energy Section of the I.C.T.P. in Trieste, and
of K.-H.R. at the Erwin Schr\"odinger Institute in Vienna.
We thank all these institutions for their hospitality and support.
N.N. and I.T. were partially supported by the Research Training Network of
the European Commission under contract MRTN-CT-2004-00514 and by the
Bulgarian National Council for Scientific Research under contract PH-1406.
}


\vskip5mm

\begin{thebibliography}{00}
\bibitem{BN06}
{\sc B. Bakalov, N.M. Nikolov,}
 Jacobi identity for vertex algebras in higher dimensions,
 J. Math. Phys. {\bf 47} (2006) 053505; math-ph/0604069.
\bibitem{BNRT07}
 {\sc B. Bakalov, N.M. Nikolov, K.--H. Rehren, I. Todorov,}
 Unitary positive-energy representations of scalar bilocal quantum fields,
 Commun. Math. Phys. {\bf 271} (2007) 223--246; math-ph/0604069.
\bibitem{BT77} {\sc V. Bargmann, I.T. Todorov,}
 Spaces of analytic functions on a complex cone as carriers for the
 symmetric tensor representations of $SO(N)$,
 J. Math. Phys. {\bf 18} (1977) 1141--1148.
\bibitem{B60}
 {\sc H.--J. Borchers,} \"Uber die Mannigfaltigkeit der
 interpolierenden Felder zu einer interpolierenden $S$-Matrix,
 N. Cim. {\bf 15} (1960) 784--794.

\bibitem{BMT88} {\sc D. Buchholz, G. Mack, I.T. Todorov,} 
 The current algebra on the circle as a germ of local field theories,
 Nucl. Phys. B (Proc. Suppl.) {\bf 5B} (1988) 20--56. 
\bibitem{DMPPT}
 {\sc V.K.~Dobrev, G.~Mack, V.B.~Petkova, S.G.~Petrova, I.T.~Todorov,}
 {\it Harmonic Analysis of the $n$-Dimensional
 Lorentz Group and Its Applications to Conformal Quantum Field
 Theory}, Springer, Berlin et al. 1977.
\bibitem{DO01}
 {\sc F.A. Dolan, H. Osborn,}
 Conformal four point functions and operator product expansion,
 Nucl. Phys. {\bf B 599} (2001) 459--496; hep-th/0011040.
\bibitem{DR03}
 {\sc M. D\"utsch, K.--H. Rehren,}
 Generalized free fields and the AdS-CFT correspondence,
 Ann. H. Poincar\'e {\bf 4} (2003) 613--635; math-ph/0209035.
\bibitem{DS63}
 {\sc N. Dunford, J. Schwartz,}
 \textit{Linear Operators, Part 2.
 Spectral Theory. Self Adjoint Operators in Hilbert Space},
 Interscience Publishers, N.Y., London, 1963.
\bibitem{M77}
 {\sc G. Mack,}
 All unitary representations of the conformal group $SU(2,2)$ with
 positive energy,
 Commun. Math. Phys. {\bf 55} (1977) 1--28.
\bibitem{NRT05}
 {\sc N.M. Nikolov, K.--H. Rehren, I.T. Todorov,}
 Partial wave expansion and Wightman positivity in conformal field theory,
 Nucl. Phys. {\bf B 722} (2005) 266--296; hep-th/0504146.
\bibitem{NST02}
 {\sc N.M. Nikolov, Ya.S. Stanev, I.T. Todorov,}
 Four dimensional CFT models with rational correlation functions,
 J. Phys. {\bf A 35} (2002) 2985--3007; hep-th/0110230.
\bibitem{NST03}
 {\sc N.M. Nikolov, Ya.S. Stanev, I.T. Todorov,}
 Globally conformal invariant gauge field theory with rational
 correlation functions,
 Nucl. Phys. {\bf B 670} (2003) 373--400; hep-th/0305200.
\bibitem{NT01}
 {\sc N.M. Nikolov, I.T. Todorov,}
 Rationality of conformally invariant local correlation functions on
 compactified Minkowsi space,
 Commun. Math. Phys. {\bf 218} (2001) 417--436; hep-th/0009004.
\bibitem{SS74} {\sc B. Schroer, J.A. Swieca,} 
 Conformal transformations of quantized fields, 
 Phys. Rev. {\bf D 10} (1974) 480--485.
\bibitem{SSV75} {\sc B. Schroer, J.A. Swieca, A.H. V\"olkel,} 
 Global operator expansions in conformally invariant relativistic
 quantum field theory,  
 Phys. Rev. {\bf D 11} (1975) 1509--1520.
\bibitem{SW}
 {\sc R.F. Streater, A.S. Wightman}, {\it PCT, Spin and
 Statistics, and All That},
 Benjamin, 1964; Princeton Univ.\ Press, Princeton, N.J., 2000.
\bibitem{T06}
{\sc I. Todorov,}
 Vertex algebras and conformal field theory models in four dimensions,
 Fortschr. Phys. {\bf 54} (2006) 496--504.
\bibitem{NRT07}
 {\sc N.M. Nikolov, K.--H. Rehren, I.T. Todorov,}
 Pole structure and biharmonic fields in conformal QFT in four
 dimensions. e-print arXiv:0711.0628, to appear in: ``LT7: Lie Theory
 and its Applications in Physics'', Proceedings Varna 2007,
 ed.\ V. Dobrev (Heron Press, Sofia).

\end{thebibliography}
\end{document}